%% file: main.tex
\documentclass[ALICE,manyauthors]{cernphprep}
\usepackage[comma,square,numbers,sort&compress]{natbib}
\usepackage{hyperref}
\usepackage{lineno}
\usepackage{xspace}
\usepackage{xcolor}
\usepackage{upgreek}
\usepackage[T1]{fontenc}
\usepackage{orcidlink}

\begin{document}
\input{commands.tex}

\begin{titlepage}
\PHyear{2024}       
\PHnumber{227}      
\PHdate{04 Sep}  

\title{J/$\psi$-hadron correlations at midrapidity in pp collisions at \s = 13 \TeV}
\ShortTitle{J/$\psi$-hadron correlations at midrapidity in pp collisions at \s = 13 \TeV}   

\Collaboration{ALICE Collaboration\thanks{See Appendix~\ref{app:collab} for the list of collaboration members}}
\ShortAuthor{ALICE Collaboration} 

\begin{abstract}
    We report on the measurement of inclusive, non-prompt, and prompt J/$\psi$-hadron correlations by the ALICE Collaboration at the CERN Large Hadron Collider in pp collisions at a center-of-mass energy of $13~\text{TeV}$. 
    The correlations are studied at midrapidity ($|y| < 0.9$) in the transverse momentum ranges $\pt < 40~\text{GeV}/c$ for the \jpsi and $0.15 < \pt < 10~\text{GeV}/c$ and $|\eta|<0.9$ for the associated hadrons. 
    The measurement is based on minimum bias and high multiplicity data samples corresponding to integrated luminosities of $L_{\text{int}} = 34~\text{nb}^{-1}$ and $L_{\text{int}} = 6.9~\text{pb}^{-1}$, respectively. 
    In addition, two more data samples are employed, requiring, on top of the minimum bias condition, a threshold on the tower energy of $E = 4$ and $9~\text{GeV}$ in the ALICE electromagnetic calorimeters, which correspond to integrated luminosities of $L_{\text{int}} = 0.9~\text{pb}^{-1}$ and $L_{\text{int}} = 8.4~\text{pb}^{-1}$
    , respectively. 
    The azimuthally integrated near and away side yields of associated charged hadrons per \jpsi trigger are presented as a function of the \jpsi and associated hadron transverse momentum. The measurements are discussed in comparison to PYTHIA calculations. 
\end{abstract}
\end{titlepage}

\setcounter{page}{2} 


\input{introduction}

\input{setup}

\input{analysis}

\input{uncertainties}

\input{results}

\input{conclusions}


\newenvironment{acknowledgement}{\relax}{\relax}
\begin{acknowledgement}
\section*{Acknowledgements}
\input{fa_2024-08-21_Opt_C_1.tex}
\end{acknowledgement}

\bibliographystyle{utphys}   
\bibliography{bibliography}

\newpage
\appendix

%
%

\section{The ALICE Collaboration}
\label{app:collab}
\input{Alice_Authorlist_2024-08-21_Opt_C_1.tex}  
\end{document}

%% file: commands.tex
%

\newcommand{\pp}           {pp\xspace}
\newcommand{\ppbar}        {\mbox{$\mathrm {p\overline{p}}$}\xspace}
\newcommand{\XeXe}         {\mbox{Xe--Xe}\xspace}
\newcommand{\PbPb}         {\mbox{Pb--Pb}\xspace}
\newcommand{\pA}           {\mbox{pA}\xspace}
\newcommand{\pPb}          {\mbox{p--Pb}\xspace}
\newcommand{\AuAu}         {\mbox{Au--Au}\xspace}
\newcommand{\dAu}          {\mbox{d--Au}\xspace}

\newcommand{\s}            {\ensuremath{\sqrt{s}}\xspace}
\newcommand{\snn}          {\ensuremath{\sqrt{s_{\mathrm{NN}}}}\xspace}
\newcommand{\pt}           {\ensuremath{p_{\rm T}}\xspace}
\newcommand{\pttrig}          {\ensuremath{p_{\rm T}^{\rm \jpsi}}\xspace}
\newcommand{\ptassoc}           {\ensuremath{p_{\rm T}^{\rm h}}\xspace}
\newcommand{\meanpt}       {$\langle p_{\mathrm{T}}\rangle$\xspace}
\newcommand{\minv}         {\ensuremath{m_{\rm inv.}}\xspace}
\newcommand{\mee}         {\ensuremath{m_{\rm e^+e^-}}\xspace}
\newcommand{\ycms}         {\ensuremath{y_{\rm CMS}}\xspace}
\newcommand{\ylab}         {\ensuremath{y_{\rm lab}}\xspace}
\newcommand{\etarange}[1]  {\mbox{$\left | \eta \right |~<~#1$}}
\newcommand{\yrange}[1]    {\mbox{$\left | y \right |~<~#1$}}
\newcommand{\dndy}         {\ensuremath{\mathrm{d}N_\mathrm{ch}/\mathrm{d}y}\xspace}
\newcommand{\dndeta}       {\ensuremath{\mathrm{d}N_\mathrm{ch}/\mathrm{d}\eta}\xspace}
\newcommand{\avdndeta}     {\ensuremath{\langle\dndeta\rangle}\xspace}
\newcommand{\dNdy}         {\ensuremath{\mathrm{d}N_\mathrm{ch}/\mathrm{d}y}\xspace}
\newcommand{\Npart}        {\ensuremath{N_\mathrm{part}}\xspace}
\newcommand{\Ncoll}        {\ensuremath{N_\mathrm{coll}}\xspace}
\newcommand{\dEdx}         {\ensuremath{\textrm{d}E/\textrm{d}x}\xspace}
\newcommand{\RpPb}         {\ensuremath{R_{\rm pPb}}\xspace}
\newcommand{\mpt}          {\ensuremath{\langle p_{\rm T}\rangle}\xspace}
\newcommand{\mptsquared}   {\ensuremath{\langle p^2_{\rm T}\rangle}\xspace}

\newcommand{\nineH}        {$\sqrt{s}~=~0.9$~Te\kern-.1emV\xspace}
\newcommand{\seven}        {$\sqrt{s}~=~7$~Te\kern-.1emV\xspace}
\newcommand{\twoH}         {$\sqrt{s}~=~0.2$~Te\kern-.1emV\xspace}
\newcommand{\twosevensix}  {$\sqrt{s}~=~2.76$~Te\kern-.1emV\xspace}
\newcommand{\five}         {$\sqrt{s}~=~5.02$~Te\kern-.1emV\xspace}
\newcommand{\thirteen}     {$\sqrt{s}~=~13$~Te\kern-.1emV\xspace}
\newcommand{\twosevensixnn}{$\sqrt{s_{\mathrm{NN}}}~=~2.76$~Te\kern-.1emV\xspace}
\newcommand{\fivenn}       {$\sqrt{s_{\mathrm{NN}}}~=~5.02$~Te\kern-.1emV\xspace}
\newcommand{\LT}           {L{\'e}vy-Tsallis\xspace}
\newcommand{\GeVc}         {Ge\kern-.1emV/$c$\xspace}
\newcommand{\MeVc}         {Me\kern-.1emV/$c$\xspace}
\newcommand{\TeV}          {Te\kern-.1emV\xspace}
\newcommand{\GeV}          {Ge\kern-.1emV\xspace}
\newcommand{\MeV}          {Me\kern-.1emV\xspace}
\newcommand{\GeVmass}      {Ge\kern-.2emV/$c^2$\xspace}
\newcommand{\MeVmass}      {Me\kern-.2emV/$c^2$\xspace}
\newcommand{\lumi}         {\ensuremath{\mathcal{L}}\xspace}

\newcommand{\ITS}          {\rm{ITS}\xspace}
\newcommand{\TOF}          {\rm{TOF}\xspace}
\newcommand{\ZDC}          {\rm{ZDC}\xspace}
\newcommand{\ZDCs}         {\rm{ZDCs}\xspace}
\newcommand{\ZNA}          {\rm{ZNA}\xspace}
\newcommand{\ZNC}          {\rm{ZNC}\xspace}
\newcommand{\SPD}          {\rm{SPD}\xspace}
\newcommand{\SDD}          {\rm{SDD}\xspace}
\newcommand{\SSD}          {\rm{SSD}\xspace}
\newcommand{\TPC}          {\rm{TPC}\xspace}
\newcommand{\TRD}          {\rm{TRD}\xspace}
\newcommand{\VZERO}        {\rm{V0}\xspace}
\newcommand{\VZEROA}       {\rm{V0A}\xspace}
\newcommand{\VZEROC}       {\rm{V0C}\xspace}
\newcommand{\Vdecay} 	   {\ensuremath{V^{0}}\xspace}

\newcommand{\jpsi}         {\ensuremath{\text{J}/\psi}\xspace}
\newcommand{\psiprime}     {\ensuremath{\psi(2\text{S})}\xspace}
\newcommand{\ccbar}        {\ensuremath{\text{c}\overline{\text{c}}}\xspace}
\newcommand{\ee}           {\ensuremath{\text{e}^{+}\text{e}^{-}}\xspace} 
\newcommand{\pip}          {\ensuremath{\pi^{+}}\xspace}
\newcommand{\pim}          {\ensuremath{\pi^{-}}\xspace}
\newcommand{\kap}          {\ensuremath{\rm{K}^{+}}\xspace}
\newcommand{\kam}          {\ensuremath{\rm{K}^{-}}\xspace}
\newcommand{\pbar}         {\ensuremath{\rm\overline{p}}\xspace}
\newcommand{\kzero}        {\ensuremath{{\rm K}^{0}_{\rm{S}}}\xspace}
\newcommand{\lmb}          {\ensuremath{\Lambda}\xspace}
\newcommand{\almb}         {\ensuremath{\overline{\Lambda}}\xspace}
\newcommand{\Om}           {\ensuremath{\Omega^-}\xspace}
\newcommand{\Mo}           {\ensuremath{\overline{\Omega}^+}\xspace}
\newcommand{\X}            {\ensuremath{\Xi^-}\xspace}
\newcommand{\Ix}           {\ensuremath{\overline{\Xi}^+}\xspace}
\newcommand{\Xis}          {\ensuremath{\Xi^{\pm}}\xspace}
\newcommand{\Oms}          {\ensuremath{\Omega^{\pm}}\xspace}
\newcommand{\degree}       {\ensuremath{^{\rm o}}\xspace}

\newcommand{\note}[1]{{\color{red}\textbf{#1}}}
\newcommand{\dphi}         {\ensuremath{\Delta\varphi}\xspace}
\newcommand{\cfphi}         {\ensuremath{C(\Delta\varphi)}\xspace}
\newcommand{\fb}         {\ensuremath{f_{\rm{B}}}\xspace}

%% file: introduction.tex
\section{Introduction}
\label{sec:introduction}

Quarkonia, bound states of charm-anticharm or beauty-antibeauty pairs, are a robust system to test our understanding of quantum chromodynamics (QCD)~\cite{Brambilla:2010cs, ALICE:2022wpn}. Quarkonium physics involves several different momentum or energy scales, the heavy-quark mass, the momentum of the quarks, as well as their kinetic energy in the bound state rest frame. While the production of the heavy-quark pair can be addressed with perturbative QCD, the bound state formation proceeds via a non-perturbative process and is subject to intense research, as summarized in Ref.~\cite{Chapon:2020heu}. For the production in pp~collisions, a few different approaches for the transition from the heavy-quark pair to the bound state are used, like the color evaporation model~\cite{Ma:2016exq,Lansberg:2020rft}, the color singlet model~\cite{Lansberg:2011hi}, and the effective field theory NRQCD~\cite{Butenschoen:2012qr,Chao:2012iv,Ma:2010jj}. Conceptually, these approaches differ in their view of the bound state formation. 
In particular at low transverse momentum, observables such as production cross sections and polarization measurements have not been sufficient to find a consensus on the most appropriate theoretical description. The measurement of associated production with other hadrons can provide additional information to discriminate between different scenarios.

The production of quarkonium in hadronic collisions is of particular interest to study this complex process in a surrounding environment of strongly interacting degrees of freedom.  
In pp~collisions at the LHC, the production of a heavy quark pair requires a large momentum transfer, which is still very small compared to the collision energy. Hence, multiple partonic interactions with a similar momentum transfer as the one required for heavy-quark production may be present in the same hadronic collision. Therefore, the measurement of heavy quarkonium as a function of charged hadrons multiplicity contains information about the abundance of such multiple partonic interactions at these energy scales. 
Previous measurements of the \jpsi production at midrapidity as a function of the event activity revealed a faster than linear increase of the \jpsi yields as a function of the charged-particle multiplicity~\cite{Acharya:2020pit}.
These measurements were described qualitatively by several model calculations implementing different physics mechanisms~\cite{Acharya:2020pit}.

In this context, the measurement of associated hadron yields to \jpsi can provide more differential information to discriminate between different quarkonium production scenarios~\cite{Lansberg:2019adr}. In high-multiplicity events, it can shed more light on the underlying mechanisms leading to the increase of the \jpsi yields with
multiplicity. This type of measurement for \jpsi was pioneered by UA1~\cite{Albajar:1987ke} and STAR~\cite{Abelev:2009qaa} collaborations as a complementary observable for the study of \jpsi~production. Similar correlation measurements were previously reported by ALICE for D mesons in pp collisions at \s = 5~\cite{ALICE:2019oyn} and 13~TeV~\cite{ALICE:2021kpy}, and for heavy-flavour electrons in pp and \pPb collisions at \snn = 5~TeV~\cite{ALICE:2023kjg}. Additionally, \jpsi production in association with charged jets was studied via the jet fragmentation functions into \jpsi in pp collisions at \s=13 TeV by the LHCb and CMS collaborations~\cite{Aaij:2017fak,Sirunyan:2019vlp,CMS:2021puf}. Together, the fragmentation and the correlation functions, discussed in this work, provide important differential information on the \jpsi hadronization phenomenology.

In this publication, the first measurement of the \jpsi-hadron correlation functions at the LHC in pp collisions is reported using the ALICE Run 2 data. The correlated yields are presented for trigger \jpsi mesons in the \pt range $\pt^{\jpsi}<40$~\GeVc and hadron \pt in the range $0.15<\ptassoc<10$~\GeVc. In addition, the correlated yields are measured in a sample of high-multiplicity events recorded using a special trigger for the \jpsi \pt range of $\pt^{\jpsi}<15$~\GeVc. The correlated yields are reported separately for inclusive, prompt, and non-prompt \jpsi. The prompt \jpsi are defined as the \jpsi originating from short-lived sources that cannot be reconstructed separately from the primary collision vertex. This includes typically directly produced \jpsi and feed-down from higher mass charmonium states. The non-prompt \jpsi are products of the weak decay of beauty hadrons, and can be experimentally separated from the prompt ones via the measurement of their secondary vertex~\cite{ALICE:2021edd}.

%% file: setup.tex
\section{The ALICE detector, dataset, and event selection}
\label{sec:setup}

The \jpsi-hadron correlation functions are constructed using \jpsi mesons and charged hadrons, which are both registered with the ALICE central barrel detectors. The \jpsi are reconstructed using the dielectron decay channel.  
A comprehensive description of the ALICE detector and its performance is available in Refs.~\cite{Aamodt:2008zz, Abelev:2014ffa}.

All the central barrel detectors are placed in a uniform magnetic field of $B = 0.5~\text{T}$ oriented along the beam direction and produced by the L3 solenoid magnet.
The track reconstruction is performed using the Inner Tracking System (ITS)~\cite{Aamodt:2010aa} and Time Projection Chamber (TPC)~\cite{Alme:2010ke} detectors.
The ITS is a cylindrical silicon detector with layers of different designs: the two innermost layers are high-resolution silicon pixel detectors (SPD), followed by two layers of silicon drift (SDD) and two layers of silicon strip (SSD) detectors. It is placed around the beam pipe at radii between $3.9$ and $43.0~\text{cm}$ and provides full coverage in azimuth. The highly segmented SPD layers provide the spatial resolution required for the reconstruction of the secondary vertices related to the weak decays of beauty hadrons, including, in particular, those decaying via \jpsi decay channels. 
The TPC, a cylindrical gaseous detector, covers the radial distance between $85$ and $250~\text{cm}$ and extends to $250~\text{cm}$ in longitudinal direction on each side of the nominal interaction point. It provides full coverage in azimuth and, combined with the ITS, a pseudorapidity range of $|\eta| < 0.9$. In addition, the TPC provides a measurement of the track specific energy loss (\dEdx) in the active gas, which is used for electron identification. 
The ElectroMagnetic Calorimeter (EMCal)~\cite{ALICE:2022qhn} and the Di-Jet Calorimeter (DCAL)~\cite{Allen:2010stl} are used for triggering on high-energy electrons and for particle identification of the matched ITS-TPC tracks by using the ratio of the calorimeter cluster energy to the matched track momentum. The 
EMCal covers a pseudorapidity interval of $|\eta| < 0.7$ and a range of $80^{\circ} < \varphi < 187^{\circ}$ in azimuth while the DCal covers a pseudorapidity range of $0.22 < |\eta| < 0.7$ in the azimuthal interval $260^{\circ} < \varphi < 320^{\circ}$ and $|\eta| < 0.7$ for $320^{\circ} < \varphi < 327^{\circ}$.
Both calorimeters have an identical design, granularity, and energy resolution and will be referred to together as EMCal throughout this paper.

In addition to the central barrel detectors, the V0 detector~\cite{Abbas:2013taa}, composed of two scintillator arrays placed on either side of the interaction point at $-90$ and $+340~\text{cm}$ and covering the pseudorapidity ranges $-3.7 < \eta < -1.7$ and $2.8 < \eta < 5.1$, respectively, is used for event triggering. Together with the SPD detector, the V0 is used to reject background from beam-gas and pileup collisions.

Multiple event triggers are employed in this analysis: minimum bias (MB), defined as the coincidence of signals in both V0 counters; high multiplicity (HM), corresponding to the $0.1\%$ events with the highest multiplicity in the V0 detector; and an EMCal trigger, which selects on the sum of energy in a sliding window of $4 \times 4$ towers above a given threshold (a tower is the smallest segmentation of the EMCal).
The EMCal trigger is activated by two different conditions, one with a threshold energy of $9~\text{GeV}$ (EG1DG1), which inspects the entire available luminosity, and one with a $4~\text{GeV}$ threshold (EG2DG2), which was downscaled by roughly a factor 10 with respect to EG1DG1. 
Both the EMCal and HM triggers include the MB trigger condition in their definition.
The analyzed samples include all data recorded by ALICE during the LHC Run 2 in 2016, 2017, and 2018 for pp collisions at \thirteen.
The maximum interaction rate for the dataset was 260~kHz, with a maximum pileup probability of 0.5$\times$10$^{-3}$.

To ensure uniform detector acceptance, events selected for analysis are required to have a reconstructed vertex within $|z_{\rm vtx}|<10~\text{cm}$ of the nominal interaction point.
Offline timing cuts based on information from the V0 detector are used to reject beam-gas events and pileup collisions occurring within the readout time of the SPD.
Pileup collisions occurring within the same LHC collision bunch crossing are rejected using offline algorithms, which identify multiple vertices~\cite{Abelev:2014ffa}. The remaining fraction of pileup events surviving the selections is negligible for the MB and EMCal triggered samples and consist of at most 2\% for the HM triggered sample~\cite{Acharya:2020pit}. However, the pileup events are not expected to affect the \jpsi-hadron correlation function since pairs of \jpsi triggers and associated hadrons from different events are not correlated and only contribute to the baseline of the correlation function.
The number of selected events and the corresponding integrated luminosity~\cite{ALICE-PUBLIC-2016-002} for the analyzed trigger samples are quoted in Table~\ref{tab:luminosities}.

The available datasets allowed the measurement of the \jpsi-hadron correlation function for \jpsi triggers in the rapidity range $|y|<0.9$ and \pt ranges of $0 < \pt < 8$~\GeVc for MB, $0 < \pt < 15$~\GeVc for HM, $8 < \pt < 15$~\GeVc for EG2DG2, and $15 < \pt < 40$~\GeVc for EG1DG1 triggers. For the case of the EMCal triggered events, the \jpsi \pt intervals are selected such that both the EG1DG1 and the EG2DG2 triggers are fully efficient for \jpsi in the selected \pt ranges. The event sample acquired using the HM trigger enables the study of the correlation function in pp collisions with a significantly larger charged-particle density than the ones selected by the MB and EMCal triggers. The average charged particle multiplicity at midrapidity for the MB dataset is 6.46~$\pm$~0.19~\cite{ALICE:2015qqj}, while for the HM data set it is approximately 4 times larger~\cite{Acharya:2020pit}. 

\begin{table}[htbp]
    \centering
    \caption{Number of selected events and corresponding integrated luminosity for the analyzed triggered samples.}
    \begin{tabular}{c | c c c c}
        
        Trigger & MB & HM & EG1DG1 & EG2DG2 \\ 
        \hline
        Number of events & $1.97 \times 10^9$ & $1.09 \times 10^9$ & $9.83 \times 10^7$ & $1.32 \times 10^8$ \\
        Integrated luminosity & $34.0 \pm 1.7~\text{nb}^{-1}$ & $6.9 \pm 0.4~\text{pb}^{-1}$ & $8.4 \pm 0.4~\text{pb}^{-1}$ & $0.9 \pm 0.1~\text{pb}^{-1}$\\
        
    \end{tabular}
    \label{tab:luminosities}
\end{table}

%% file: analysis.tex
\section{Data analysis}
\label{sec:analysis}

The correlation function of \jpsi (trigger) and charged hadrons (associated particles) are obtained by reconstructing \jpsi mesons using the dielectron channel. It closely follows the procedure employed for the measurement of the \jpsi production cross section at $\sqrt{s} = 5$ and $13~\text{TeV}$, published by ALICE in Refs.~\cite{Acharya:2019lkw,ALICE:2021dtt,ALICE:2021edd}. The prompt and non-prompt \jpsi are separated by reconstructing the displaced secondary vertex of the decaying beauty hadron. The azimuthal correlation functions are constructed using the difference in azimuth between the trigger dielectron and associated charged hadrons, as described below.

\subsection{Electron and hadron track selection}
\label{sec:trackSelection}

The electron candidates are good quality tracks selected in the central barrel acceptance ($|\eta| < 0.9$), with a minimum \pt of $1~\text{GeV}/c$ and matched between ITS and TPC. The selection criteria follow those used in previous analyses reported in Refs.~\cite{Acharya:2019lkw,ALICE:2021dtt,ALICE:2021edd}. 
Electrons are identified using the TPC \dEdx measurement and are required to be within a band of $[-1.5, 3]~\sigma$ relative to the expectation for electrons, where $\sigma$ is the \dEdx resolution associated with each track. 
Proton and pion contamination is minimized by rejecting tracks compatible with the proton or pion \dEdx expectation within $3.5\sigma$. 
In the EMCal-triggered data sample, at least one of the electrons of the \jpsi candidate pair is required to have a matched cluster in the EMCal. 
If a cluster match is found, the pion rejection is replaced by a selection on $E/p \in [0.8, 1.3]$, where $E$ is the energy of the cluster matched to the track and $p$ is the track momentum. 
This calorimeter-based particle identification improves the electron selection efficiency at large momentum where the TPC pion rejection would otherwise remove most of the candidate electron tracks.

The associated charged hadrons are tracks reconstructed in both the ITS and TPC selected in an $\eta$ range of $|\eta| < 0.9$ and with a minimum \pt of 150~\MeVc. A distance-of-closest approach (DCA) to the primary vertex lower than $1~\text{cm}$ in the transverse direction and $3~\text{cm}$ along the beam axis is imposed. 
These requirements select a sample of tracks containing less than $20\%$ contamination from secondary particles, falling with increasing transverse momentum. The contamination with secondaries in the measured correlated yields is taken into account via the efficiency correction computed from a Monte Carlo (MC) simulation reweighted to match the fraction of secondary tracks observed in data.

\subsection{\jpsi yield measurement}
\label{sec:signalExtraction}

\begin{figure}[htbp]
    \centering
    \includegraphics[width=.45\textwidth]{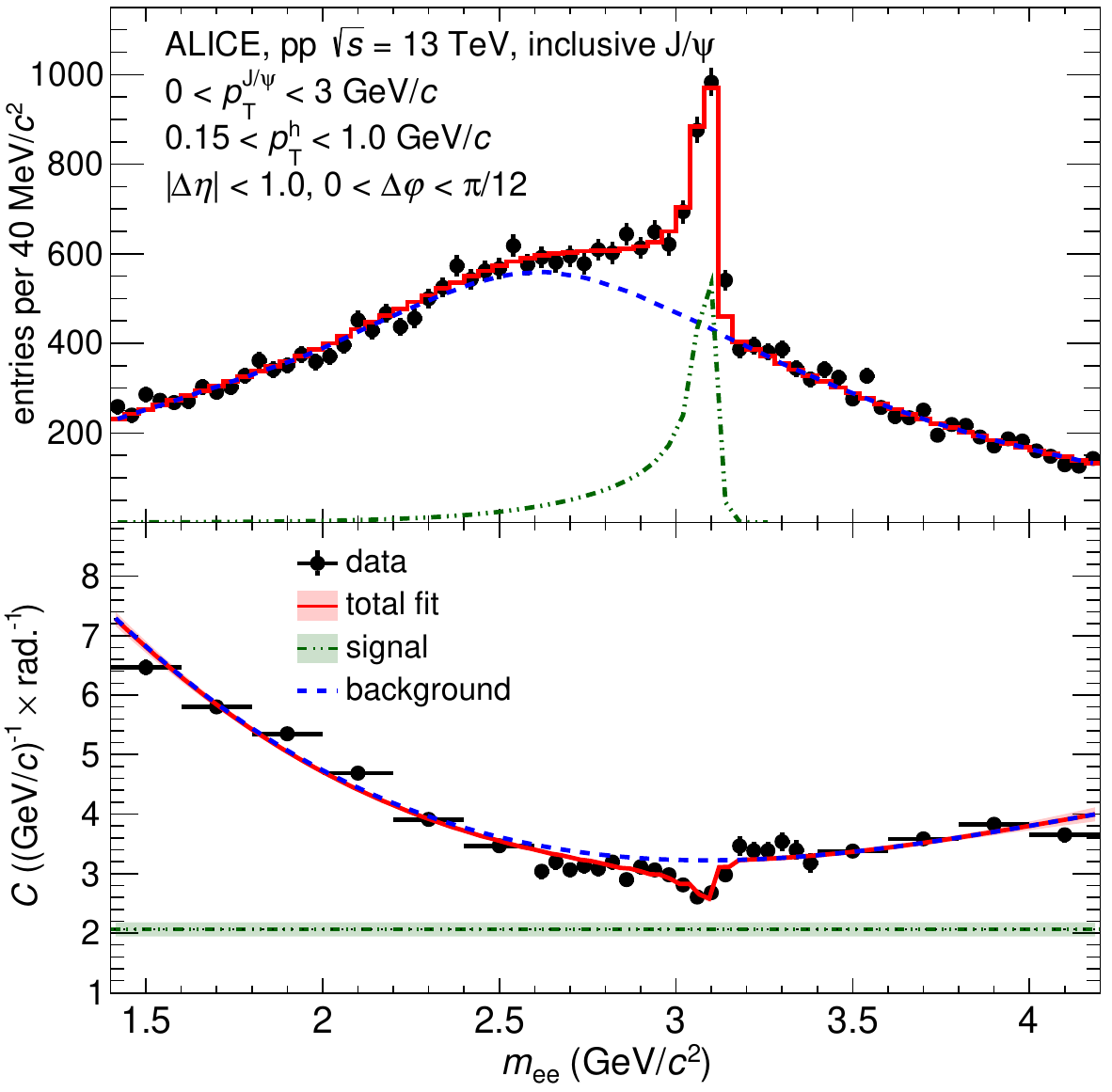}
    \includegraphics[width=.45\textwidth]{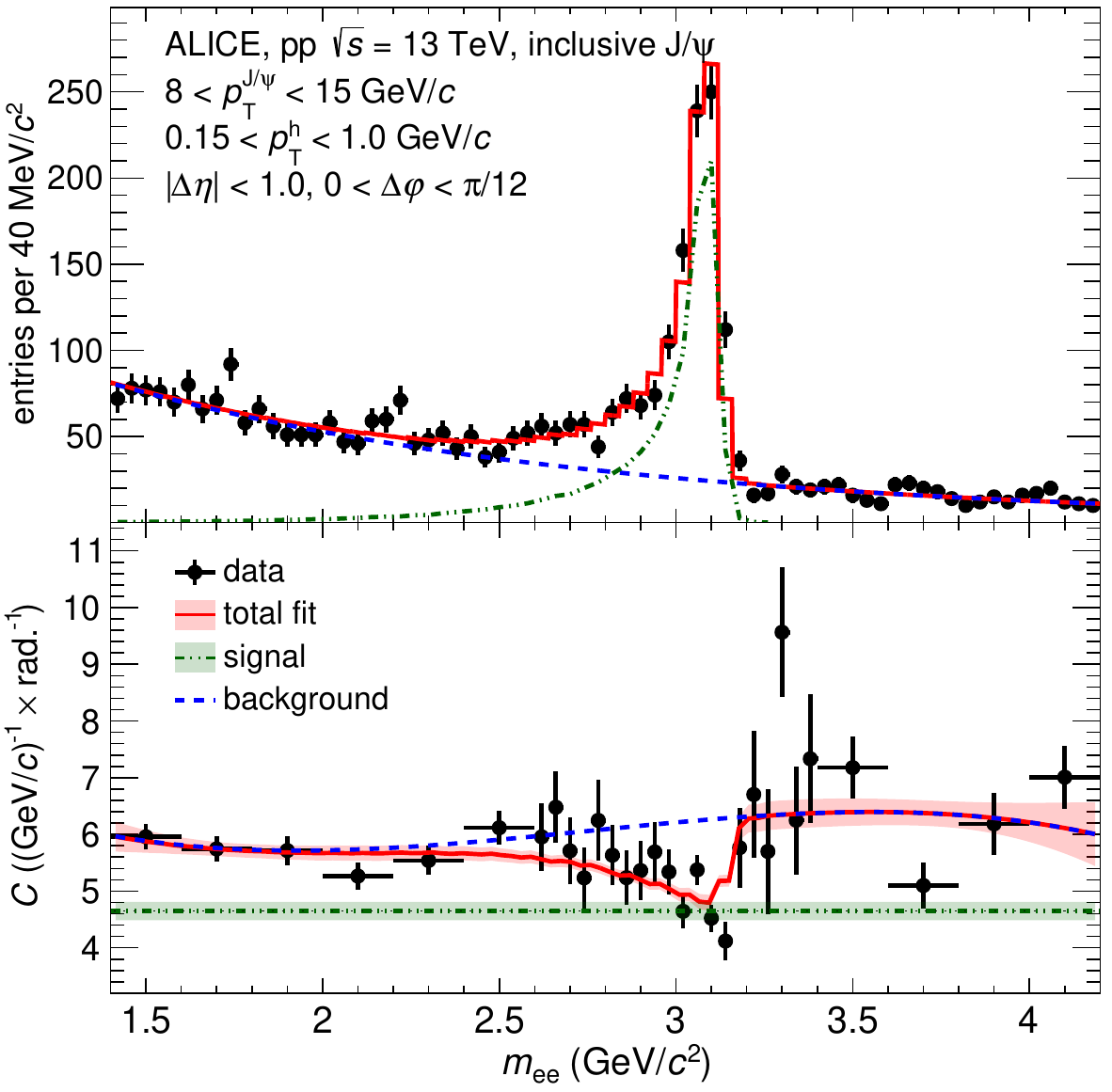}
    \caption{Examples of the extraction of the per trigger correlated charged hadron yield ($0.15<\ptassoc<1.0$~\GeVc) for inclusive \jpsi in the \pt ranges $0<\pt^{\jpsi}<3$~\GeVc (left) and $8<\pt^{\jpsi}<15$~\GeVc (right) for $|\Delta\eta|<1$ and $0<\Delta\varphi<\pi/12$. The top panels show the invariant mass distribution of the dielectrons with the signal and background components as described in Sec.~\ref{sec:signalExtraction}. The bottom panels show the invariant mass dependence of the charged hadrons correlated yield. The total fit function ($C_{\mee}$), the signal ($C_{\jpsi}$) and the total background ($C_{\rm bkg}(\mee)$) components as defined in Eq.~\ref{eq:correlation} are also shown. The band around the total fit function in the bottom panels shows the 1$\sigma$ confidence interval.}
    \label{fig:correlation_signal_extraction}
\end{figure}

Dielectrons are built from all possible opposite-sign~(OS) electron pairs constructed using the selected electron tracks within the same event~(SE).
The \jpsi signal is extracted from the invariant mass distribution of these pairs by subtracting the combinatorial and correlated backgrounds.
The combinatorial background, pairs of electrons originating from uncorrelated processes, is estimated using the event-mixing~(ME), where dielectrons are constructed using electrons from different events.
The mixing is done in event pools with events belonging to the same data-taking run (a period of continuous data taking) and with a similar vertex position. Since an arbitrary number of ME pairs can be obtained with this technique, the ME invariant mass distribution is normalized using the ratio between the like-sign (LS) pairs from the SE to the one from the ME distributions, as described in Ref.~\cite{ALICE:2021dtt}.  
The correlated background, originating mainly from semi-leptonic decays of heavy-flavor hadrons~\cite{Acharya:2018ohw}, is determined by means of a fit of the invariant mass distribution after subtraction of the combinatorial background.
This fit includes a template obtained from a MC simulation, described below, for the \jpsi signal and an empirical shape for the correlated background parameterized using a second-order polynomial function for pair $\pt<3$~\GeVc and an exponential function for $\pt>3$~\GeVc.
For high transverse momentum pairs, $\pt>5$~\GeVc, the invariant mass distribution is fitted directly, without subtraction of the combinatorial background. 
The upper panels of Fig.~\ref{fig:correlation_signal_extraction} illustrate the signal extraction procedure for inclusive \jpsi candidates in a low (left panel) and a high (right panel) transverse momentum interval. For a given invariant mass interval, the \jpsi signal yield is obtained by subtracting the background contribution estimated from the fitting procedure.

\subsection{\jpsi - hadron correlation functions}
\label{sec:correlation}

For a sample of electron-pairs in a transverse momentum interval \pttrig and invariant mass \mee, the correlation function $C$ with hadrons in a transverse momentum interval \ptassoc is calculated as 
\begin{equation}
C({\rm \Delta}\eta,{\rm \Delta}\varphi;\pttrig,\mee,\ptassoc) = \frac{1}{N_{\rm trig}} \cdot \frac{S({\rm \Delta}\eta,{\rm \Delta}\varphi)}{B({\rm \Delta}\eta,{\rm \Delta}\varphi)} \cdot B(0,0)
\label{eq:corrDef}
\end{equation}
where $N_{\rm trig}$ is the number of \jpsi triggers, while ${\rm \Delta}\eta$ and ${\rm \Delta}\varphi$ are the difference in pseudorapidity and azimuthal angle, respectively, between the dielectron trigger and associated hadron. The $S$ and $B$ are the doubly-differential distributions of \jpsi-hadron pairs obtained using trigger \jpsi and associated hadrons from the same event or different events, respectively. The distribution obtained by combining \jpsi and hadrons from different events (also known as the mixed-event procedure) is used to correct for geometrical acceptance effects and is normalized by its amplitude at ${\rm \Delta}\eta = {\rm \Delta}\varphi = 0$, B(0,0), where the acceptance for \jpsi-hadron pairs is maximal.
In the mixed-event procedure, trigger dielectrons from one event are mixed with associated hadrons from a different event with similar event vertex position and close in time to ensure similar detector conditions. For the \jpsi candidates reconstructed in EMCal-triggered events, the event mixing is performed using associated hadrons from MB-triggered events. This is necessary to avoid fake correlations originating from trivial geometrical acceptance effects due to the azimuthal coverage of the EMCal detector, which selects only events that contain a trigger \jpsi in the relatively narrow EMCal acceptance and recoils in the opposite direction.

The dielectron-pair distributions are corrected for acceptance and reconstruction efficiency by applying a weight to each dielectron-hadron pair, which is the product of the corrections corresponding to the candidate \jpsi and to the associated hadron. The number of \jpsi triggers, $N_{\rm trig}$, is also corrected for the dielectron acceptance and reconstruction efficiency which means that this correction largely cancels for the correlation function. However, since the \jpsi efficiency depends on \pt, especially for large \pttrig intervals, this leads to a small correction of the correlation functions. The correction corresponding to the associated hadrons takes into account the reconstruction efficiency of the primary charged hadrons, originating from the collision, and the contamination by secondary tracks originating from long-lived weak decays and secondary interactions in the detector material. The hadron reconstruction efficiency, which relies on a MC simulation, ranges between 85\% and 90\% for $\pt>0.5$~\GeVc and drops to approximately 70\% below 0.5~\GeVc. Since the relative contribution of secondary hadrons is not properly simulated in the MC sample, a data driven procedure is used, based on extracting the fraction of secondaries from data and combining it with the primary and secondary-hadron efficiencies computed using the MC sample. The \pt-dependent fraction of secondary hadrons is evaluated from fits to the measured distributions of the DCA transverse projection for the selected charged hadrons with MC-generated templates corresponding to primary and secondary tracks. The fraction of secondaries in the selected associated hadron sample ranges from approximately 20\% at the lowest hadron \pt and goes down to approximately 3\% at $\pt=10$~\GeVc.

In any \pttrig, \ptassoc and ${\rm \Delta}\varphi$ interval, the correlation function is built for a given invariant mass, $C(\mee)$, obtained as defined by Eq.~\ref{eq:corrDef} and averaged in the interval $|\Delta\eta|<1$, $C(\mee)$. It is a weighted average of the contributions from \jpsi, $C_{\jpsi}$ (independent of \mee), and from background dielectrons, $C_{\rm bkg}(\mee)$,
\begin{equation}
    C(\mee) = f(\mee) \times C_{\jpsi} + (1-f(\mee)) \times C_{\rm bkg}(\mee),
    \label{eq:correlation}
\end{equation}
with $f$ being the fraction of \jpsi among the total number of dielectrons with invariant mass \mee. The extraction of the correlation component $C_{\jpsi}$ relies on fitting Eq.~\ref{eq:correlation} to the invariant mass dependent associated yield $C(\mee)$ obtained for each combination of \pttrig, \ptassoc and $\Delta\varphi$ intervals. The background correlation component $C_{\rm bkg}(\mee)$ is parameterized using a fifth-order polynomial for $\pttrig<3$~\GeVc and a third-order polynomial otherwise. The signal fraction $f(\mee)$ is obtained from the fit of the invariant mass distribution of dielectron pairs in the transverse momentum interval corresponding to \pttrig, as explained in Sec.~\ref{sec:signalExtraction} and illustrated in the upper panels of Figure~\ref{fig:correlation_signal_extraction}. The fit of the associated yield is performed in the invariant mass interval $1.4<\mee<4.2$~\GeVmass and is illustrated in the lower panels of Fig.~\ref{fig:correlation_signal_extraction} for inclusive \jpsi in the \pttrig ranges $\pttrig<3$~\GeVc (left) and $8<\pttrig<15$~\GeVc (right).

\subsection{Prompt and non-prompt \jpsi separation}
\label{sec:promptSeparation}

\begin{figure}[htbp]
    \centering
    \includegraphics[width=.45\textwidth]{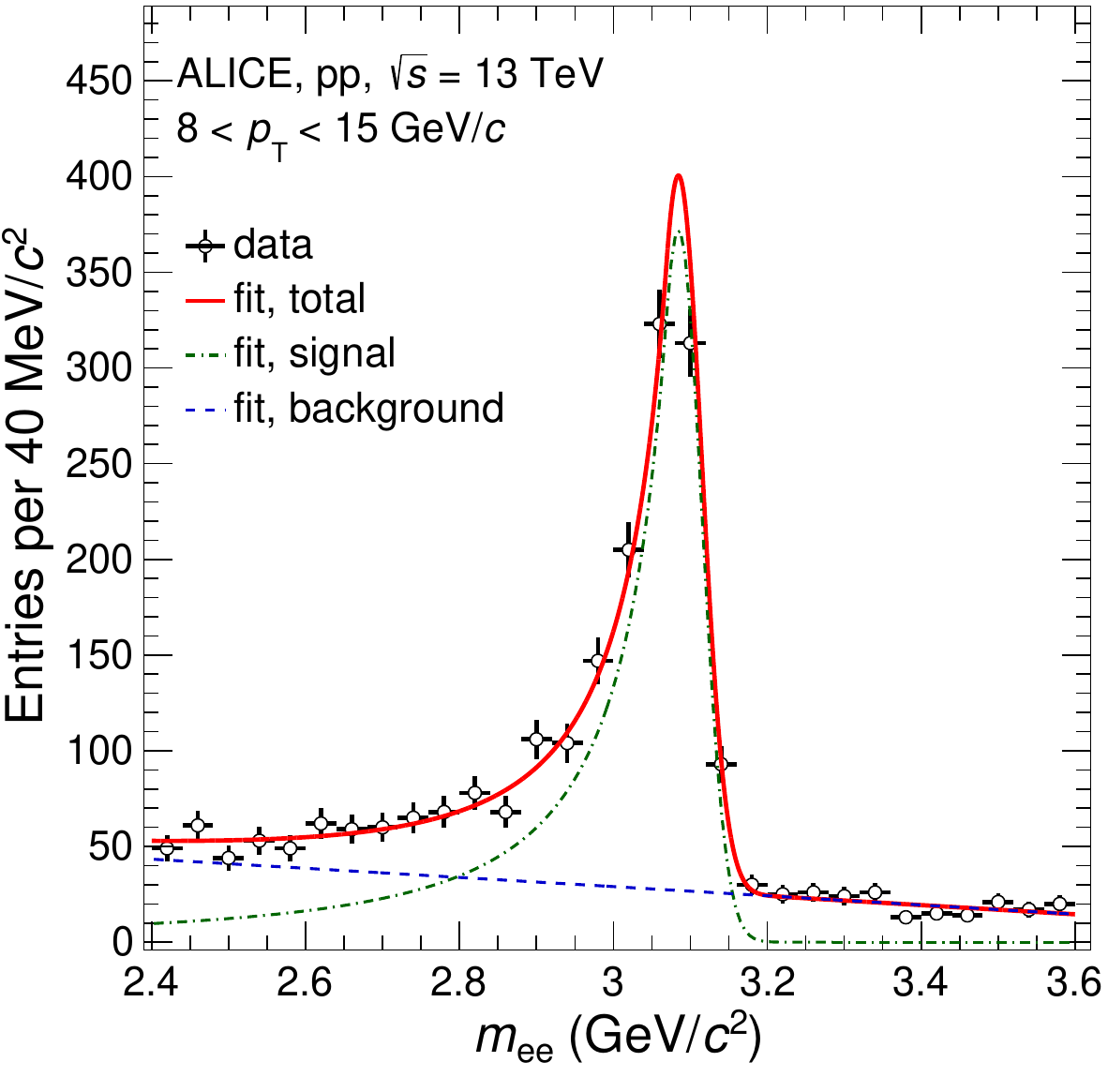}
    \includegraphics[width=.45\textwidth]{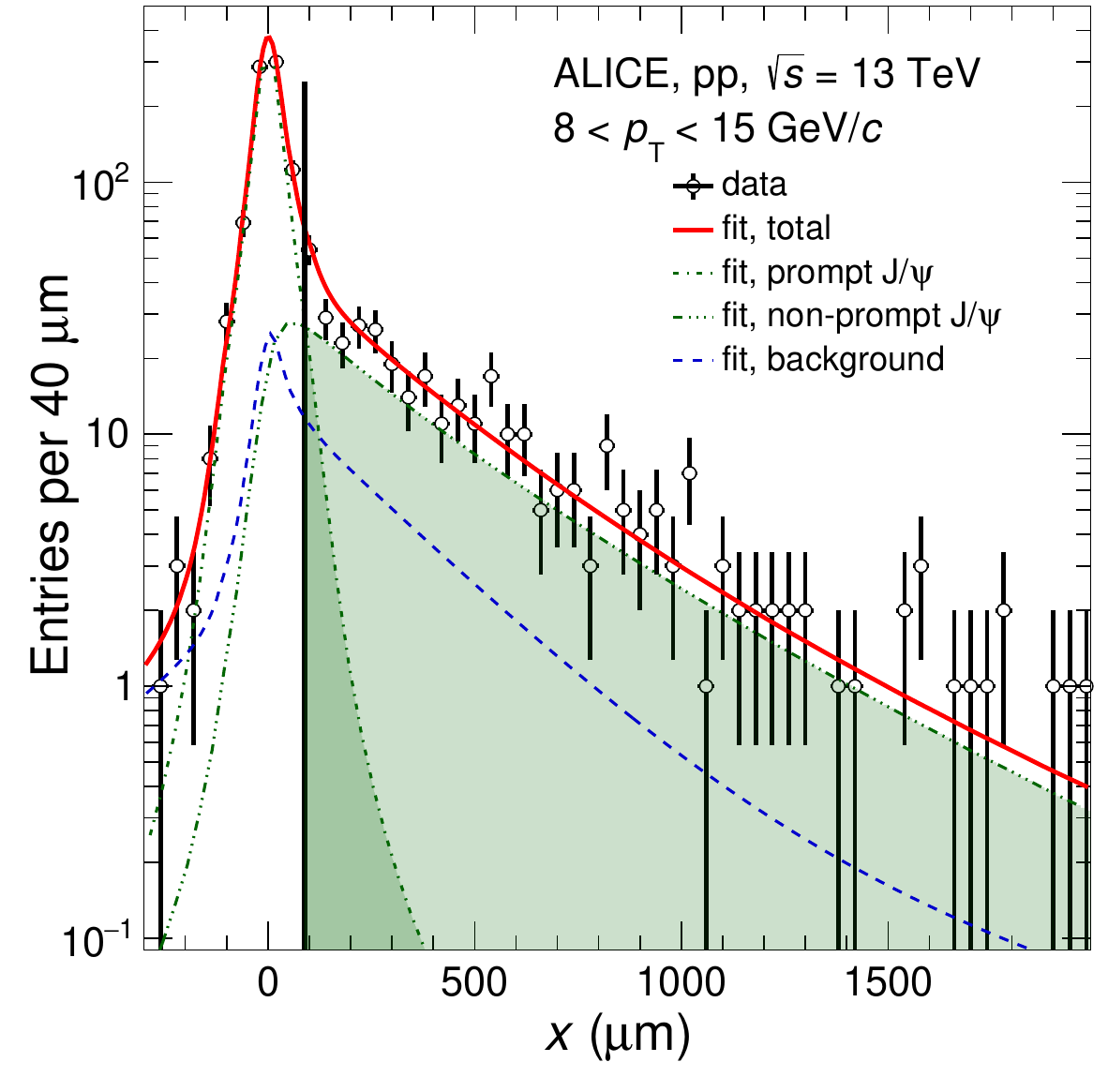}
    \caption{Simultaneous fit of the dielectron invariant mass and pseudo-proper decay length two-dimensional distribution in the $8<\pt^{\jpsi}<15$~\GeVc interval. Left: projection on the invariant mass axis. Right: Projection on the pseudo-proper decay length axis. The various components of the fit are also shown. The vertical black line at $x\approx$100~${\upmu}$m shows the lower cut value selecting a non-prompt \jpsi sample with less than 10\% contamination from prompt \jpsi. 
    }
    \label{fig:decaylength_distributions}
\end{figure}

Prompt and non-prompt \jpsi are separated based on the pseudo-proper decay length ($x$) defined as 
\begin{equation}
    x = L_{\rm xy} \cdot \frac{c \cdot m^{\jpsi}}{\pt^{\jpsi}},
    \label{eq:x}
\end{equation}
where $L_{\rm xy}$ is the transverse projection of the distance between the secondary vertex (constructed using the two electrons) and the primary vertex of the collision.
Small values of $x$ distributed symmetrically around zero, with a spread determined by the ITS spatial resolution of~$\approx$ 50~${\upmu}$m, are characteristic of prompt \jpsi, while large positive $x$ values are characteristic of non-prompt \jpsi originating in the weak decay of beauty hadrons.
ALICE measured the non-prompt \jpsi cross section in proton-proton collisions at $\s = 13~\text{TeV}$, as reported in Ref.~\cite{ALICE:2021edd}, where the different shapes of the pseudo-proper decay length distributions of prompt and non-prompt \jpsi were exploited.
For a given \jpsi transverse momentum interval \pttrig, the correlation function for inclusive \jpsi, determined using Eq.~\ref{eq:correlation}, can be described with two components, corresponding to the prompt and non-prompt \jpsi 
\begin{equation}
    C_{\rm inclusive}(\Delta\varphi) = f_{\rm B}(\pttrig) \times C_{\jpsi \leftarrow h_{\rm B}}(\Delta\varphi) + (1-f_{\rm B}(\pttrig)) \times C_{\rm prompt}(\Delta\varphi),
    \label{eq:corrPrompt}
\end{equation}
where $f_{\rm B}(\pttrig)$ is the fraction of non-prompt \jpsi mesons in the transverse momentum interval corresponding to \pttrig. 
Since $f_{\rm B}$ is an increasing function of \pt, leading to significant variations within the \pttrig intervals employed in this analysis, an interpolation procedure is used to obtain a continuous dependence of $f_{\rm B}$ on \pttrig. The interpolation is done using a fit of existing measurements at midrapidity in pp collisions by ALICE at \s = 5, 7 and 13~TeV~\cite{Abelev:2012gx, ALICE:2021edd}, CMS at \s = 5 and 7~TeV~\cite{CMS:2017exb, CMS:2010nis} and ATLAS at \s = 7 and 8~TeV~\cite{ATLAS:2015zdw}. The measurements at midrapidity in p$\bar{\rm p}$ collisions at \s = 1.96~TeV by CDF~\cite{CDF:2004jtw}, and those at forward-rapidity by LHCb in pp collisions at \s = 13~TeV~\cite{LHCb:2015foc} were also used. In order to account for the energy or rapidity dependence of the \pt-dependent \fb ratios from the various datasets, each measured \fb distribution is scaled by a constant factor, which is obtained by fitting the ratio of the ALICE measurement at 13~TeV to the given measurement in the overlapping \pt interval. The resulting set of experimental data points is fitted with a universal curve, which is obtained as the ratio between the non-prompt \jpsi cross section obtained using a FONLL calculation~\cite{Cacciari2012,Cacciari2015} and a power-law function for the inclusive \jpsi cross section, as described in Ref.~\cite{Bossu2011}. Systematic uncertainties due to the intrinsic scales of the FONLL calculation and residual variations of the \pt dependence with energy or rapidity are calculated and reported in the last row of Table~\ref{tab:systematics}. The largest uncertainty contributions originate from the FONLL calculations and the scaling of the forward LHCb measurement. 

A two-step approach is used in order to extract the prompt and non-prompt \jpsi-hadron correlation functions. 
First, the correlation function for non-prompt \jpsi is obtained using the procedure described in Sec.~\ref{sec:correlation}, but applying a lower threshold on $x$ such that a high purity non-prompt \jpsi sample is selected. The threshold value is determined based on templates of the prompt and non-prompt \jpsi $x$ distributions, with a relative normalization corresponding to the $f_{B}(\pttrig)$ value obtained from the fit procedure explained above, and allowing a maximum of 10\% contamination with prompt \jpsi. The template shapes depend mainly on the \jpsi \pt and on the number of pixel hits in the SPD layers, which drive the spatial resolution of the secondary vertexing.
The templates, as described in detail in Ref.~\cite{ALICE:2021edd}, 
are obtained based on a MC simulation tuned to match the vertexing resolution observed in data. An illustration of these, corresponding to the \pt interval $8<\pttrig<15$~\GeVc, is shown in Fig.~\ref{fig:decaylength_distributions}, where these templates, together with a background template obtained as explained in Ref.~\cite{ALICE:2021edd,Abelev:2012gx}, are used to fit the $x$ distribution seen in data. The low-$x$ threshold corresponding to a non-prompt \jpsi sample with a 10\% contamination with prompt \jpsi is indicated with a vertical black line. 
Second, the prompt \jpsi correlation function is estimated using Eq.~\ref{eq:corrPrompt}, once the inclusive and non-prompt \jpsi components are known.

\subsection{Monte Carlo simulations}
\label{sec:mc}
In order to compute efficiency corrections for both the \jpsi triggers and the associated hadrons, and to construct the prompt and non-prompt \jpsi pseudoproper decay length templates, two sets of Monte Carlo simulations are employed. The first one, used to compute the associated-hadron efficiency, is a sample of minimum bias pp collisions simulated with PYTHIA8 and the Monash 2013 tune~\cite{Skands:2014pea}. The second one is a set of minimum bias pp collisions simulated with PYTHIA6 and the Perugia 2011 tune~\cite{Skands:2010ak} with injected unpolarized prompt and non-prompt \jpsi. The non-prompt \jpsi kinematics is generated using the Perugia 2011 PYTHIA tune, while for the prompt \jpsi, a spectrum based on a phenomenological interpolation of
measurements at RHIC, CDF, and the LHC~\cite{Bossu2011} and a uniform distribution in rapidity is used. The \jpsi are forced to decay via the dielectron channel using the EvtGen package~\cite{LANGE2001152} with the PHOTOS model~\cite{BARBERIO1994291}. The generated particles are transported through a detailed simulation of the ALICE apparatus using GEANT3~\cite{Brun:1082634}. Detector responses and calibrations in MC simulations are tuned to data and consider the
time-dependent conditions of all detectors included in the data acquisition. In particular, 
the track parameters from the simulations are tuned to minimize the residual discrepancy of the DCA distributions between data and MC, as described in Ref.~\cite{ALICE:2018szk}.

%% file: uncertainties.tex
\section{Systematic uncertainties}
\label{sec:uncertainties}

\begin{table}[htbp]
    \centering
    \caption{Summary of the systematic uncertainties on the correlated-hadron yields from all the sources mentioned in the text. The degree of correlation over $\Delta\varphi$, trigger-\pt and associated-hadron-\pt is also quoted. Uncertainties are provided as ranges to keep the table compact.}
    \begin{tabular}{c | c | c c c}
        
        Source & Uncertainty & \multicolumn{3}{c}{Correlation} \\ 
        \hline
         & & $\Delta\varphi$ & trigger \pt & associated \pt \\
        \hline
        Charged-hadron efficiency & 2--3\% & correlated & correlated & correlated \\
        Electron tracking & negligible & correlated & correlated & correlated \\
        Electron identification & 1--10\% & correlated & uncorrelated & correlated \\
        Correlated yield fit & 1--9\% & uncorrelated & uncorrelated & uncorrelated \\
        prompt \jpsi contamination & 0--4\% & correlated & uncorrelated & correlated \\
        $f_{\rm B}$ & 1--8\% & correlated & correlated & correlated
        
    \end{tabular}
    \label{tab:systematics}
\end{table}

Several sources of systematic uncertainty affecting the \jpsi correlation functions are considered, including the reconstruction efficiency of the \jpsi and charged hadrons, the fraction of \jpsi signal ($f(\mee)$) used in Eq.~\ref{eq:correlation}, the parameterization of the background correlation function $C_{\rm bkg}(\mee)$, the contamination with prompt \jpsi in the selected non-prompt \jpsi sample, and the fraction of \jpsi originating from beauty hadron decays ($f_{B}$). 
The impact from the non-zero \jpsi polarization measured by the LHC~\cite{ALICE:2011gej, LHCb:2013izl} and Tevatron~\cite{CDF:2007msx} experiments is negligible. This is mainly due to the fact that the correlated charged hadron yields are reported per \jpsi trigger, and to the fact that the \jpsi efficiency variations due to the polarization scenarios allowed by the existing measurements do not alter the \jpsi distribution in any of the considered \jpsi \pt intervals. A summary with all the systematic uncertainties is given in Table~\ref{tab:systematics} and in the following we provide a detailed description. 

The systematic uncertainty on the associated hadron yield ranges between 2\% and 3\% and is dominated by the uncertainties on the ITS-TPC track matching efficiency, with a smaller contribution originating from the contamination with secondary particles. The uncertainty on the ITS-TPC matching is determined based on residual differences observed between data and MC simulations~\cite{ALICE:2017olh}. The uncertainty due to the contamination with secondary particles is estimated by alternatively using the longitudinal DCA component for extracting the fraction of secondaries, instead of the standard approach employing the transverse DCA. This uncertainty is considered fully correlated over ${\rm \Delta}\varphi$, \pttrig and \ptassoc.
The uncertainty due to the electron tracking is negligible, mainly due to the per-trigger normalization. Residual effects are checked by repeating the analysis with variations of the electron candidate track quality criteria and are found to be negligible.

The uncertainty due to the electron identification is studied by repeating the analysis using different electron PID selection criteria, which have a large impact on both the \jpsi efficiency and the fraction of \jpsi signal, $f(\mee)$. Both the TPC and EMCal electron selection criteria are varied, leading to systematic variations ranging between 1 and 10\%. The impact of the \jpsi signal and background invariant mass description on $f(\mee)$ are negligible as shown in Ref.~\cite{ALICE:2021dtt}. This is due to the very good description of the \jpsi signal shape in the Monte-Carlo simulations and of the usage of the event mixing procedure which closely matches the combinatorial background. For a given \jpsi \pt interval, these uncertainties are considered correlated over $\Delta\varphi$ and associated-hadron \pt and fully uncorrelated between the different \jpsi \pt intervals.

The uncertainty related to the fit of the invariant mass dependent correlated yield, as described by Eq.~\ref{eq:correlation}, is studied by varying both the lower and upper limits of the mass range in which the fit is performed by 200~\MeVmass. 
The assigned systematic uncertainty, which varies from 1 to 9\%, is computed as the difference between the maximum and minimum value of the correlation yields divided by $\sqrt{12}$, corresponding to the ratio of the standard deviation and the full spread of a uniform distribution. It is considered to be uncorrelated for all the studied kinematic intervals. The large number of associated yield extractions required made impractical the use of a functional form other than a polynomial to parameterize the background correlation component, $C_{\rm bkg}(\mee)$. Therefore, the choice of the polynomial function order was done such that the fit properly converges for all the considered cases. It is then considered that the systematic uncertainty obtained based on the fit range variations and the yield statistical uncertainty provide a good estimation of the total uncertainty on the correlated yield extraction.

Since a completely pure sample of non-prompt \jpsi cannot be obtained with the available datasets,  residual contamination is always present and considered as a systematic analysis uncertainty. This is estimated by modifying the pseudoproper decay length selection to allow a smaller (5\%) and a larger (20\%) prompt \jpsi contamination in the non-prompt sample. The difference obtained between the two cases, divided by $\sqrt{12}$, is taken as a systematic uncertainty, ranging from 0 to 4\%, and is considered to be uncorrelated over the trigger \pt and correlated in associated hadron \pt and $\Delta\varphi$. 

Finally, the uncertainty corresponding to the determination of the fraction $f_{\rm B}$ of \jpsi originating from beauty-hadron decays is dominated by the uncertainties of the measurements used in the $f_{\rm B}(\pt)$ interpolation procedure described in the previous section. 
This uncertainty, which varies from 1 to 8\%, is considered to be correlated between all the considered trigger and associated \pt intervals as well as $\Delta\varphi$.

%% file: results.tex
\section{Results}
\label{sec:results}

The corrected correlation functions, described in Sec.~\ref{sec:correlation}, are obtained in $\Delta\varphi$ intervals of $\pi/12$ for several \jpsi \pt intervals and three associated-hadron \pt intervals.
The main features of the correlation functions, in a first-order hard scattering di-jet production picture, are the so called near-side (NS) peak centered around $\dphi=0$, which is related to the structure of the fragmenting jet containing the trigger particle, and the away-side (AS) peak centered around $\dphi=\pi$, which is related to the fragmenting recoil jet. The near-side peak is relatively narrow for high-\pt trigger particles due to the kinematic bias made when a trigger particle is selected. In contrast, the away-side peak is typically wider due to the relatively softer particles composing the recoil jet and to jet acoplanarity, predominantly originating from initial-state and/or final-state radiation~\cite{PHENIX:2006gto}. In addition, the shape of the correlation functions is sensitive to the hadronization and particle decays which alter the kinematics of the detected jet particles relative to the parent parton.
While the peak structures of the correlation functions reflect the particle production in the hard scattering leading to the production of the \jpsi, the absolute scale of the correlation function includes also random combinatorics, which is proportional to the event multiplicity and is typically independent of \dphi. Thus, it is convenient to look at the baseline subtracted correlation functions, with the baseline $b$ determined from a fit of the correlation function with a function of the form 
\begin{equation}
    C(\Delta\varphi) = b + a_{\text{NS}} \times \text{e}^{-\frac{(\Delta\varphi)^2}{2\sigma_{\text{NS}}^2}} + a_{\text{AS}} \times \text{e}^{-\frac{(\Delta\varphi - \pi)^2}{2\sigma_{\text{AS}}^2}} ,
    \label{eq:correlation_yield_extraction}
\end{equation}
where $a_{\text{NS}}$($a_{\text{AS}}$) and $\sigma_{\text{NS}}$($\sigma_{\text{AS}}$) are the amplitudes and widths of two Gaussians, corresponding to the near and away sides as indicated by the parameter indices.

\begin{figure}[htbp]
    \centering
    \includegraphics[width=.98\textwidth]{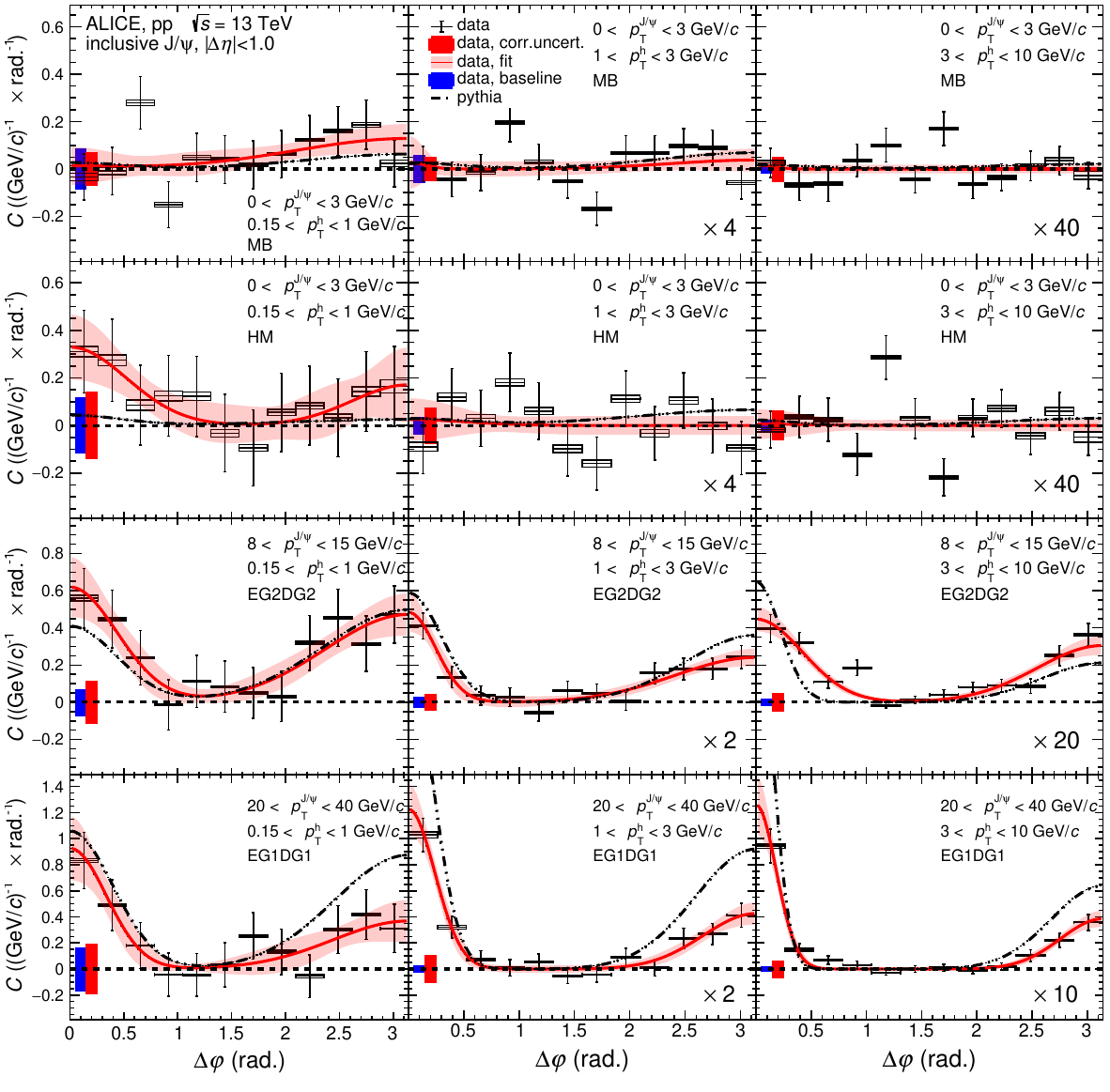}
    \caption{Baseline-subtracted inclusive \jpsi-hadron correlation functions corresponding, from top to bottom, to the MB, HM, EG2DG2, and EG1DG1 event triggers. The differential correlated yields are normalized per radian and per \GeVc. From left to right, the associated-hadron \pt intervals are $0.15<\ptassoc<1$ \GeVc, $1<\ptassoc<3$ \GeVc and $3<\ptassoc<10$~\GeVc. For better visibility, the middle and right panel correlation functions are scaled by a factor, given in the lower right corner of each panel. The error bars on the data points indicate statistical uncertainties while the boxes show the uncorrelated systematic uncertainties. The correlated systematic and the baseline uncertainties are shown as filled boxes around 0. The fit of the data using Eq.~\ref{eq:correlation_yield_extraction} is indicated by the red solid line while the band around the fit shows the 1$\sigma$ confidence interval. Calculations using PYTHIA are shown as black dash-dotted lines.}
    \label{fig:cfInclusive}
\end{figure}

A few examples of the baseline-subtracted correlation functions, \cfphi, for each hardware trigger used in this analysis are shown in Fig.~\ref{fig:cfInclusive}. The plots refer to inclusive \jpsi production. From top to bottom the rows correspond to the MB ($0<\pt^{\jpsi}<3$~\GeVc), HM ($0<\pt^{\jpsi}<3$~\GeVc), EG2DG2 ($8<\pt^{\jpsi}<15$~\GeVc) and EG1DG1 ($20<\pt^{\jpsi}<40$~\GeVc) triggers. From left to right, the panels correspond to associated-hadrons in the \pt ranges $0.15<\ptassoc<1$ \GeVc, $1<\ptassoc<3$ \GeVc and $3<\ptassoc<10$~\GeVc, respectively. 
These correlation functions exhibit the characteristic features described above, particularly for the EG2DG2 and EG1DG1 datasets, which select higher \pt \jpsi (triggers). Even though this analysis does not work on full reconstructed jets, some features can be naturally understood in the context of jets:  it can be observed that the near-side peak amplitude increases with increasing \jpsi \pt, and the peak becomes narrower~\cite{ParticleDataGroup:2022pth}, which is due to the boosted kinematics of the jet components but also to the bias imposed on the jet fragmentation by selecting a high \pt trigger particle. The away-side peak is wider compared to the near-side as mentioned above, and also displays a relatively smaller amplitude, possibly due to the rapidity swing of the recoil jet and the limited experimental acceptance. 

\begin{figure}[htbp]
    \centering
    \includegraphics[width=.99\textwidth]{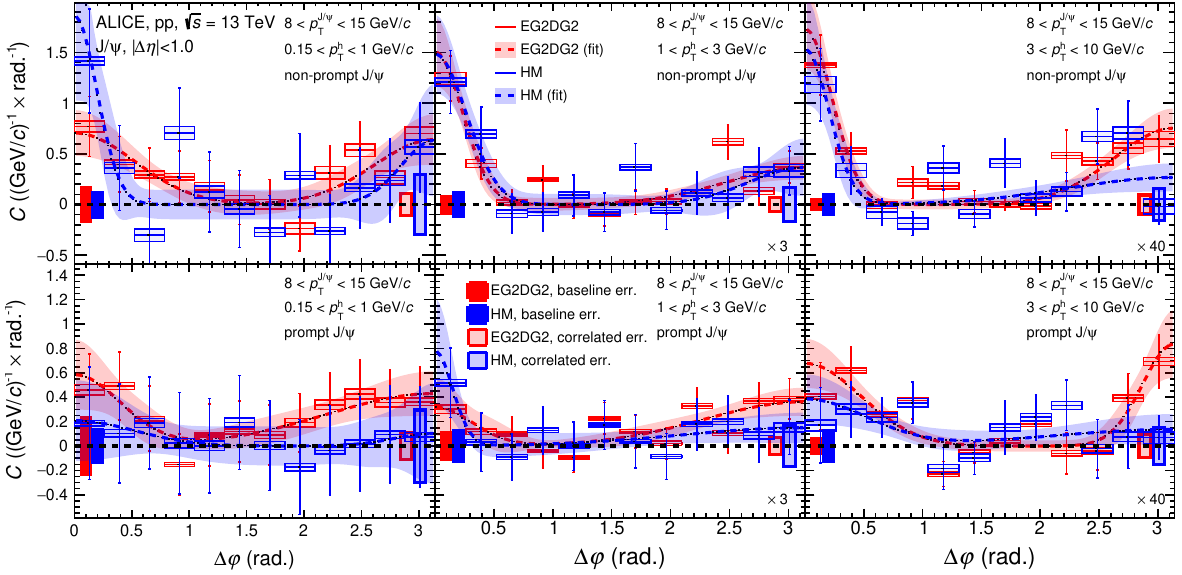}
    \caption{Baseline subtracted non-prompt (top panels) and prompt (bottom panels) \jpsi-hadron correlation functions obtained using the EG2DG2 (red), and HM (blue) triggers for \jpsi in the \pt interval $8<\pt^{\jpsi}<15$~\GeVc and associated hadrons in the \pt intervals $0.15<\ptassoc<1$, $1<\ptassoc<3$ and $3<\ptassoc<10$~\GeVc. The differential correlated yields are normalized per radian and per \GeVc. Baseline and correlated uncertainties are shown as boxes around 0, while fits of the correlation functions with Eq.~\ref{eq:correlation_yield_extraction} are shown as a dashed line with a colored band, which represent the 1$\sigma$ fit uncertainties.}
    \label{fig:cfPromptNonPrompt}
\end{figure}

The correlation functions are compared to Monte Carlo simulations of inelastic pp collisions at \mbox{$\s=13$~\TeV} using the PYTHIA 8.3 generator with the Monash 2013 tune~\cite{Skands:2014pea}. It should be noted that this PYTHIA tune employs fragmentation functions tuned on measurements from \ee and ep collisions, while hadronization can be different in \pp collisions as suggested by multiple recent experimental results~\cite{ALICE:2023sgl}. For the calculations compared to the HM triggered sample, one per mille of events with the highest charged particle multiplicity in the V0 detector acceptance were selected, analogously to the HM triggered experimental data. The calculations are shown in Figure~\ref{fig:cfInclusive} as dash-dotted black lines. In order to remove effects due to the different baselines seen in simulations, the PYTHIA correlation functions are also fitted with the Eq.~\ref{eq:correlation_yield_extraction}, and the corresponding baseline is subtracted.

Figure~\ref{fig:cfPromptNonPrompt} shows an example of a comparison of the baseline-subtracted correlation functions for non-prompt and prompt \jpsi in EG2DG2 and HM triggered events. This comparison is chosen as enough data were available for both triggers to extract the correlated yields, in the same \jpsi \pt interval, $8<\pt^{\jpsi}<15$~\GeVc. The top panels correspond to the non-prompt \jpsi, while the bottom panels correspond to the prompt \jpsi.  
Despite the significantly different baseline values, $b$, due to the much higher multiplicity seen in the HM triggers, the two baseline-subtracted correlation functions show peak structures which are compatible within uncertainties. 
Both the non-prompt (top panels) and prompt (bottom panels) \jpsi correlation functions shown in Fig.~\ref{fig:cfPromptNonPrompt} indicate correlation patterns on both the near and away sides, although for the non-prompt \jpsi the statistical significance is stronger. 

\begin{figure}[htbp]
    \centering
\includegraphics[width=.95\textwidth]{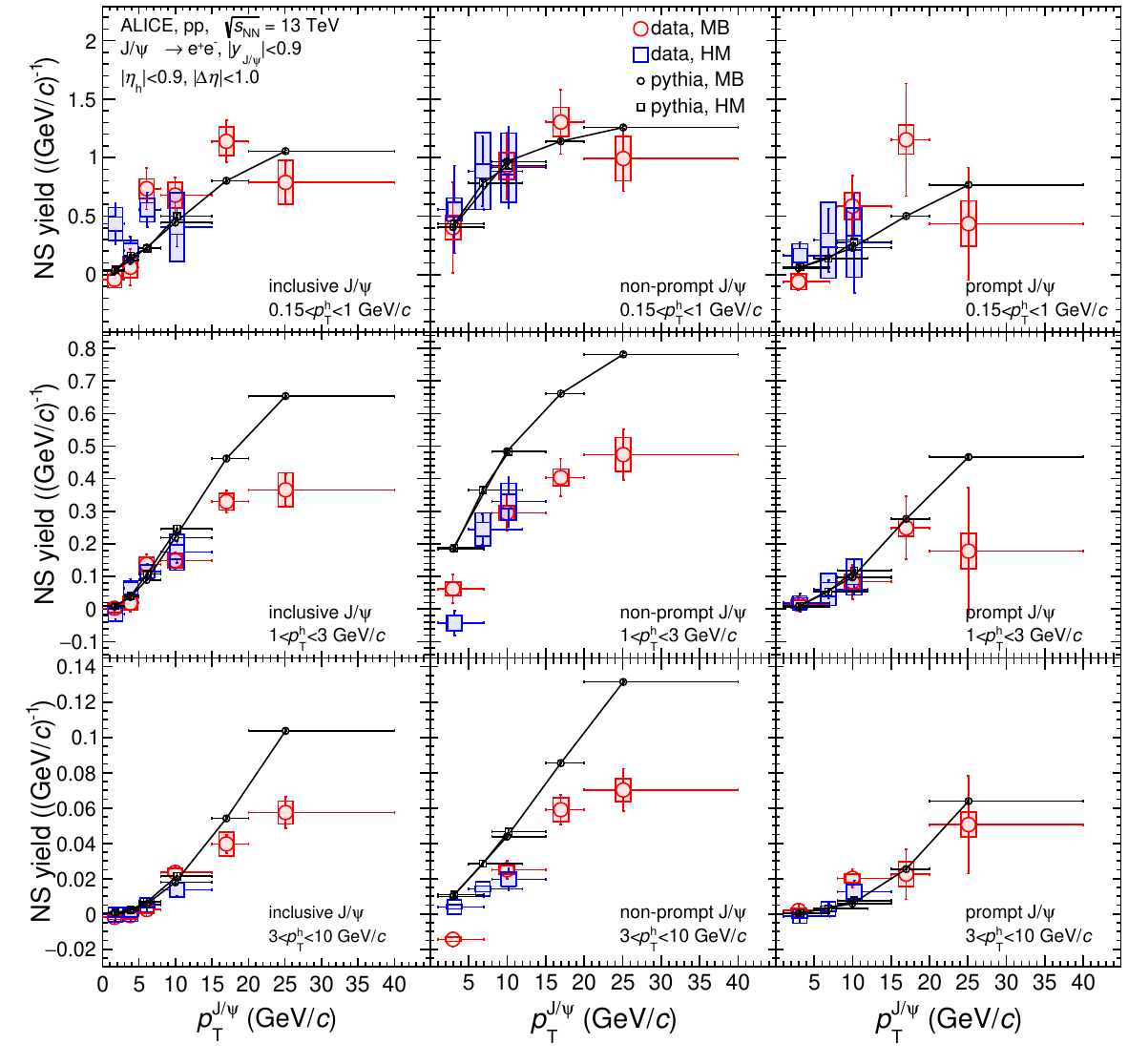}
    \caption{Near-side correlated particle yields as a function of the \jpsi trigger \pt. The left, middle, and right panels correspond to the inclusive, non-prompt, and prompt \jpsi, respectively. The top, middle, and bottom panels correspond to the three associated \pt intervals. Error bars on the data points represent statistical uncertainties, while the boxes represent systematic uncertainties. The data points are plotted at the average \jpsi \pt in each particular trigger \pt range, while the horizontal extent of the error bars indicate the \pt interval width. The red data points include the combination of the MB, EG2DG2, and EG1DG1 triggered events (named MB in the figure legend), while the blue data points represent the HM-triggered events. The data is compared to yields obtained from PYTHIA simulations, represented with solid red (MB) and dashed-blue (HM) lines.}
    \label{fig:yield_NS}
\end{figure}

\begin{figure}[htbp]
    \centering
    \includegraphics[width=.95\textwidth]{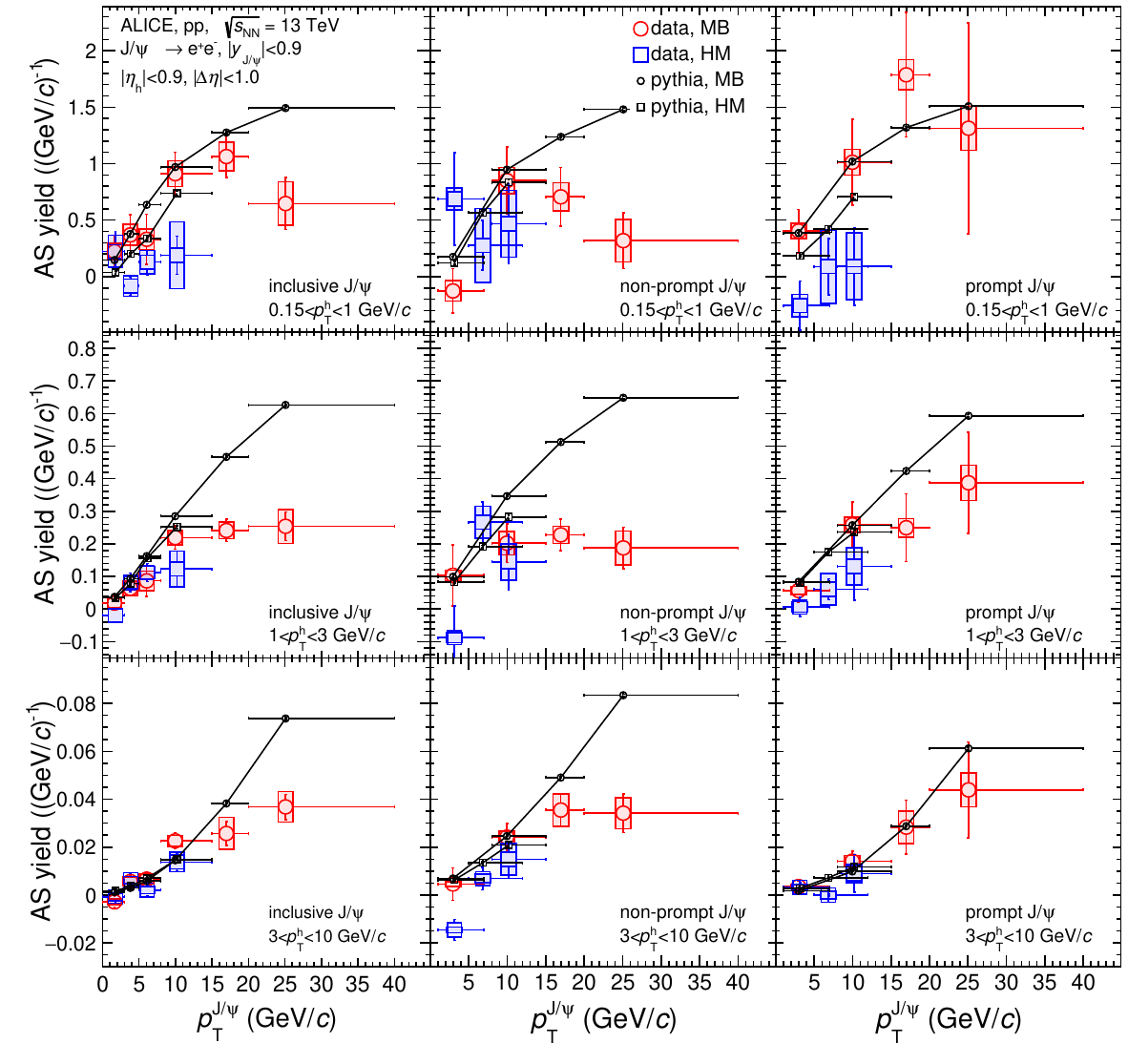}
    \caption{Away-side correlated particle yields as a function of the \jpsi trigger \pt. The left, middle, and right panels correspond to the inclusive, non-prompt, and prompt \jpsi, respectively. The top, middle, and bottom panels correspond to the three associated \pt intervals. Error bars on the data points represent statistical uncertainties, while the boxes represent systematic uncertainties. The data points are plotted at the average \jpsi \pt in each particular trigger \pt range, while the horizontal extent of the error bars indicate the \pt interval width. The red data points include the combination of the MB, EG2DG2, and EG1DG1 triggered events (named MB in the figure legend), while the blue data points represent the HM-triggered events. The data is compared to yields obtained from PYTHIA simulations, represented with solid red (MB) and dashed-blue lines (HM).}
    \label{fig:yield_AS}
\end{figure}

For a quantitative study of the charged particle production in association with \jpsi mesons, the NS and AS yields integrated over $\Delta\varphi$ are obtained from the fit of Eq.~\ref{eq:correlation_yield_extraction} to each correlation function using the statistical and uncorrelated systematic uncertainties added in quadrature. The yields are computed by integrating the corresponding NS and AS Gaussians. The statistical uncertainties on the correlated yields are also obtained using the fit parameter covariance matrix. 
As a systematic check, the yields are obtained also by using a bin counting method, as the number of counts above the baseline. In order to optimize the significance, the counting is done by using only the bins found within one $\sigma$ on either side of the NS and AS Gaussian functions. A minimum of two ${\rm \Delta}\varphi$ bins are used in the counting, in the cases where the Gaussian width is smaller than a bin width. 
The total yield is obtained after accounting for the portion of the Gaussian that covers the bins used in counting.
In the following, the reported integrated yields are the ones obtained from integrating the fitted Gaussians, while half of the difference between the fit and the counting method is taken as a systematic uncertainty. 

The near-side and away-side correlated yields obtained from data and PYTHIA simulations are shown in Figs.~\ref{fig:yield_NS} and~\ref{fig:yield_AS}, respectively, as a function of the trigger \jpsi \pt and separately for the inclusive, non-prompt and prompt \jpsi. 
The systematic uncertainties shown on the data points are dominated by the uncorrelated uncertainties discussed in Sec.~\ref{sec:uncertainties}. The combined results from the MB, EG2DG2 and EG1DG1 triggered events are shown in red and denoted as "MB" in the figure legend, while the ones from the HM-triggered events are shown in blue. For the case of prompt and non-prompt \jpsi in the HM event sample, the results shown include three \pttrig intervals which have partial overlap, namely $1 < \pt < 7$~\GeVc,  $5 < \pt < 12$~\GeVc, and $8 < \pt < 15$~\GeVc. This choice was made in order to have overlap with the corresponding results from the MB and EG2DG2 triggered event samples, while the interval $5 < \pt < 12$~\GeVc was included to provide an additional data point at an intermediate \pttrig.

For a given \ptassoc interval, the NS associated yields shown in Figure~\ref{fig:yield_NS} grow as a function of \pttrig for inclusive, prompt and non-prompt \jpsi. The trend is significant for the hadrons with $\pt>1$~\GeVc, while for the lowest \pt the measurement precision does not allow to conclude.
For \jpsi with $\pt>15$~\GeVc, the NS per-trigger yield is about 1 for hadrons in the range $0.15<\pt<1$~\GeVc, dropping to 0.3-0.4 for $1<\pt<3$~\GeVc, and to about 0.05 for $3<\pt<10$~\GeVc.
The near-side yields associated with non-prompt \jpsi are larger than those associated to prompt \jpsi, with a statistical significance of 2.2, 3.4 and 1.7 $\sigma$ for the associated \pt intervals $0.15<\ptassoc<1$~\GeVc, $1<\ptassoc<3$~\GeVc and $3<\ptassoc<10$~\GeVc, respectively. The significance evaluation includes both the statistical and uncorrelated systematic uncertainties. 
The associated yields observed in the HM triggered events and in MB / EMCal triggered events show a good agreement within uncertainties both for inclusive, non-prompt and prompt \jpsi. This is an indication that the parton fragmentation into \jpsi, prompt or non-prompt, is not significantly modified in events with high event activity relative to MB events. This is consistent with observations made in Ref.~\cite{ALICE:2021kpy} where prompt $\rm D^{0}$ meson-hadron correlations are compared across different V0M multiplicity classes, covering also the current HM event sample, and revealing no appreciable change. 
The PYTHIA calculations, done in the same kinematic intervals as the data, are generally in good qualitative agreement with NS results, however, hadron yields with $\pt>1$~\GeVc associated with non-prompt \jpsi are overestimated by the model. The fragmentation functions for non-prompt \jpsi in fully reconstructed jets reported by LHCb in Ref.~\cite{Aaij:2017fak} are in good agreement with PYTHIA. Altogether, the results might point to hadronization effects which are not well reproduced in PYTHIA.

The away-side yields, related to the recoil jet, are shown in Fig.~\ref{fig:yield_AS}. 
The PYTHIA calculations are in good agreement with the results obtained for prompt \jpsi over the entire \pttrig range, while for inclusive and in particular non-prompt \jpsi the model starts to diverge from the measurements for \pt above 15~\GeVc, overestimating the experimental yields. 
The associated yields obtained in HM triggered events are overall statistically compatible with those obtained in MB/EMCal triggered events.
However, in particular for prompt \jpsi and in the $0.15<\ptassoc<1.0$~\GeVc interval, there is a hint of lower correlated hadron yields in HM triggered events than in MB events. The statistical significance is approximately 2.3$\sigma$ and a difference in the yields for the two triggered samples is also observed in the PYTHIA calculations. As also discussed in Ref.~\cite{ALICE:2023plt}, the lower away-side yields are likely related to the definition of the HM trigger, which requires a high threshold on charged-particle multiplicity in the V0 detector acceptance. This is introducing a bias towards events where the recoil jet is in the V0 acceptance, hence lower away-side yields would be seen at midrapidity.

%% file: conclusions.tex
\section{Summary}
\label{sec:summary}

The first \jpsi-hadron correlation analysis in pp collisions at the LHC is reported using four datasets at \mbox{\s = 13~TeV} recorded with a minimum-bias ($L_{\rm int} = 34$~nb$^{-1}$), a high-multiplicity ($L_{\rm int} = 6.9$~pb$^{-1}$) and an EMCal trigger used with a threshold on the tower energy of 4 GeV ($L_{\rm int} = 0.9$~pb$^{-1}$) and 9 GeV \mbox{($L_{\rm int} = 8.4$~pb$^{-1}$)}. For a given hadron \pt range, the associated yields for inclusive, prompt and non-prompt \jpsi in both the near- and away-side grow with the increase in \pt of the trigger \jpsi. 
The yields extracted in minimum-bias or EMCal-triggered events agree well within experimental uncertainties with those obtained in high-multiplicity events, hinting that the fragmentation of the jet producing the \jpsi does not change with the event multiplicity. The PYTHIA calculations in general show a good agreement with data, but some tension is observed. In the case of non-prompt \jpsi and for hadrons with $\pt>1$~\GeVc, PYTHIA overpredicts the measurements for the near-side, while in the away-side the correlated yields are overestimated only for \jpsi \pt above 15~\GeVc. This might indicate hadronization effects in pp collisions which are not well described by PYTHIA. 
Neither the data nor PYTHIA calculations exhibit differences in correlated yields between MB and HM-triggered events, with the exception of the away-side correlated yield with hadrons of $\pt<1$~\GeVc. The lower AS correlated yields seen in high-multiplicity events are interpreted as being related to the trigger bias introduced by the HM trigger.
The measurement of the \jpsi correlated hadron yields will be improved with the datasets collected during the LHC Run 3, thanks to the increase in integrated luminosity but also to the higher precision of the separation between prompt and non-prompt \jpsi of the upgraded ALICE detector~\cite{ALICE:2023udb}.

%% file: fa_2024-08-21_Opt_C_1.tex

The ALICE Collaboration would like to thank all its engineers and technicians for their invaluable contributions to the construction of the experiment and the CERN accelerator teams for the outstanding performance of the LHC complex.
The ALICE Collaboration gratefully acknowledges the resources and support provided by all Grid centres and the Worldwide LHC Computing Grid (WLCG) collaboration.
The ALICE Collaboration acknowledges the following funding agencies for their support in building and running the ALICE detector:
A. I. Alikhanyan National Science Laboratory (Yerevan Physics Institute) Foundation (ANSL), State Committee of Science and World Federation of Scientists (WFS), Armenia;
Austrian Academy of Sciences, Austrian Science Fund (FWF): [M 2467-N36] and Nationalstiftung f\"{u}r Forschung, Technologie und Entwicklung, Austria;
Ministry of Communications and High Technologies, National Nuclear Research Center, Azerbaijan;
Conselho Nacional de Desenvolvimento Cient\'{\i}fico e Tecnol\'{o}gico (CNPq), Financiadora de Estudos e Projetos (Finep), Funda\c{c}\~{a}o de Amparo \`{a} Pesquisa do Estado de S\~{a}o Paulo (FAPESP) and Universidade Federal do Rio Grande do Sul (UFRGS), Brazil;
Bulgarian Ministry of Education and Science, within the National Roadmap for Research Infrastructures 2020-2027 (object CERN), Bulgaria;
Ministry of Education of China (MOEC) , Ministry of Science \& Technology of China (MSTC) and National Natural Science Foundation of China (NSFC), China;
Ministry of Science and Education and Croatian Science Foundation, Croatia;
Centro de Aplicaciones Tecnol\'{o}gicas y Desarrollo Nuclear (CEADEN), Cubaenerg\'{\i}a, Cuba;
Ministry of Education, Youth and Sports of the Czech Republic, Czech Republic;
The Danish Council for Independent Research | Natural Sciences, the VILLUM FONDEN and Danish National Research Foundation (DNRF), Denmark;
Helsinki Institute of Physics (HIP), Finland;
Commissariat \`{a} l'Energie Atomique (CEA) and Institut National de Physique Nucl\'{e}aire et de Physique des Particules (IN2P3) and Centre National de la Recherche Scientifique (CNRS), France;
Bundesministerium f\"{u}r Bildung und Forschung (BMBF) and GSI Helmholtzzentrum f\"{u}r Schwerionenforschung GmbH, Germany;
General Secretariat for Research and Technology, Ministry of Education, Research and Religions, Greece;
National Research, Development and Innovation Office, Hungary;
Department of Atomic Energy Government of India (DAE), Department of Science and Technology, Government of India (DST), University Grants Commission, Government of India (UGC) and Council of Scientific and Industrial Research (CSIR), India;
National Research and Innovation Agency - BRIN, Indonesia;
Istituto Nazionale di Fisica Nucleare (INFN), Italy;
Japanese Ministry of Education, Culture, Sports, Science and Technology (MEXT) and Japan Society for the Promotion of Science (JSPS) KAKENHI, Japan;
Consejo Nacional de Ciencia (CONACYT) y Tecnolog\'{i}a, through Fondo de Cooperaci\'{o}n Internacional en Ciencia y Tecnolog\'{i}a (FONCICYT) and Direcci\'{o}n General de Asuntos del Personal Academico (DGAPA), Mexico;
Nederlandse Organisatie voor Wetenschappelijk Onderzoek (NWO), Netherlands;
The Research Council of Norway, Norway;
Pontificia Universidad Cat\'{o}lica del Per\'{u}, Peru;
Ministry of Science and Higher Education, National Science Centre and WUT ID-UB, Poland;
Korea Institute of Science and Technology Information and National Research Foundation of Korea (NRF), Republic of Korea;
Ministry of Education and Scientific Research, Institute of Atomic Physics, Ministry of Research and Innovation and Institute of Atomic Physics and Universitatea Nationala de Stiinta si Tehnologie Politehnica Bucuresti, Romania;
Ministry of Education, Science, Research and Sport of the Slovak Republic, Slovakia;
National Research Foundation of South Africa, South Africa;
Swedish Research Council (VR) and Knut \& Alice Wallenberg Foundation (KAW), Sweden;
European Organization for Nuclear Research, Switzerland;
Suranaree University of Technology (SUT), National Science and Technology Development Agency (NSTDA) and National Science, Research and Innovation Fund (NSRF via PMU-B B05F650021), Thailand;
Turkish Energy, Nuclear and Mineral Research Agency (TENMAK), Turkey;
National Academy of  Sciences of Ukraine, Ukraine;
Science and Technology Facilities Council (STFC), United Kingdom;
National Science Foundation of the United States of America (NSF) and United States Department of Energy, Office of Nuclear Physics (DOE NP), United States of America.
In addition, individual groups or members have received support from:
Czech Science Foundation (grant no. 23-07499S), Czech Republic;
FORTE project, reg.\ no.\ CZ.02.01.01/00/22\_008/0004632, Czech Republic, co-funded by the European Union, Czech Republic;
European Research Council (grant no. 950692), European Union;
ICSC - Centro Nazionale di Ricerca in High Performance Computing, Big Data and Quantum Computing, European Union - NextGenerationEU;
Academy of Finland (Center of Excellence in Quark Matter) (grant nos. 346327, 346328), Finland.

%% file: Alice_Authorlist_2024-08-21_Opt_C_1.tex
\begin{flushleft} 
\small

S.~Acharya\,\orcidlink{0000-0002-9213-5329}\,$^{\rm 127}$, 
A.~Agarwal$^{\rm 135}$, 
G.~Aglieri Rinella\,\orcidlink{0000-0002-9611-3696}\,$^{\rm 32}$, 
L.~Aglietta\,\orcidlink{0009-0003-0763-6802}\,$^{\rm 24}$, 
M.~Agnello\,\orcidlink{0000-0002-0760-5075}\,$^{\rm 29}$, 
N.~Agrawal\,\orcidlink{0000-0003-0348-9836}\,$^{\rm 25}$, 
Z.~Ahammed\,\orcidlink{0000-0001-5241-7412}\,$^{\rm 135}$, 
S.~Ahmad\,\orcidlink{0000-0003-0497-5705}\,$^{\rm 15}$, 
S.U.~Ahn\,\orcidlink{0000-0001-8847-489X}\,$^{\rm 71}$, 
I.~Ahuja\,\orcidlink{0000-0002-4417-1392}\,$^{\rm 37}$
% ...

\section*{Affiliation Notes}

$^{\rm I}$ Deceased\\
$^{\rm II}$ Also at: Max-Planck-Institut fur Physik, Munich, Germany\\
$^{\rm III}$ Also at: Italian National Agency for New Technologies, Energy and Sustainable Economic Development (ENEA), Bologna, Italy\\
$^{\rm IV}$ Also at: Dipartimento DET del Politecnico di Torino, Turin, Italy\\
$^{\rm V}$ Also at: Yildiz Technical University, Istanbul, T\"{u}rkiye\\
$^{\rm VI}$ Also at: Department of Applied Physics, Aligarh Muslim University, Aligarh, India\\
$^{\rm VII}$ Also at: Institute of Theoretical Physics, University of Wroclaw, Poland\\
$^{\rm VIII}$ Also at: An institution covered by a cooperation agreement with CERN\\

\section*{Collaboration Institutes}

$^{1}$ A.I. Alikhanyan National Science Laboratory (Yerevan Physics Institute) Foundation, Yerevan, Armenia\\
$^{2}$ AGH University of Krakow, Cracow, Poland\\
$^{3}$ Bogolyubov Institute for Theoretical Physics, National Academy of Sciences of Ukraine, Kiev, Ukraine\\
$^{4}$ Bose Institute, Department of Physics  and Centre for Astroparticle Physics and Space Science (CAPSS), Kolkata, India\\
$^{5}$ California Polytechnic State University, San Luis Obispo, California, United States\\
$^{6}$ Central China Normal University, Wuhan, China\\
$^{7}$ Centro de Aplicaciones Tecnol\'{o}gicas y Desarrollo Nuclear (CEADEN), Havana, Cuba\\
$^{8}$ Centro de Investigaci\'{o}n y de Estudios Avanzados (CINVESTAV), Mexico City and M\'{e}rida, Mexico\\
$^{9}$ Chicago State University, Chicago, Illinois, United States\\
$^{10}$ China Institute of Atomic Energy, Beijing, China\\
$^{11}$ China University of Geosciences, Wuhan, China\\
$^{12}$ Chungbuk National University, Cheongju, Republic of Korea\\
$^{13}$ Comenius University Bratislava, Faculty of Mathematics, Physics and Informatics, Bratislava, Slovak Republic\\
$^{14}$ Creighton University, Omaha, Nebraska, United States\\
$^{15}$ Department of Physics, Aligarh Muslim University, Aligarh, India\\
$^{16}$ Department of Physics, Pusan National University, Pusan, Republic of Korea\\
$^{17}$ Department of Physics, Sejong University, Seoul, Republic of Korea\\
$^{18}$ Department of Physics, University of California, Berkeley, California, United States\\
$^{19}$ Department of Physics, University of Oslo, Oslo, Norway\\
$^{20}$ Department of Physics and Technology, University of Bergen, Bergen, Norway\\
$^{21}$ Dipartimento di Fisica, Universit\`{a} di Pavia, Pavia, Italy\\
$^{22}$ Dipartimento di Fisica dell'Universit\`{a} and Sezione INFN, Cagliari, Italy\\
$^{23}$ Dipartimento di Fisica dell'Universit\`{a} and Sezione INFN, Trieste, Italy\\
$^{24}$ Dipartimento di Fisica dell'Universit\`{a} and Sezione INFN, Turin, Italy\\
$^{25}$ Dipartimento di Fisica e Astronomia dell'Universit\`{a} and Sezione INFN, Bologna, Italy\\
$^{26}$ Dipartimento di Fisica e Astronomia dell'Universit\`{a} and Sezione INFN, Catania, Italy\\
$^{27}$ Dipartimento di Fisica e Astronomia dell'Universit\`{a} and Sezione INFN, Padova, Italy\\
$^{28}$ Dipartimento di Fisica `E.R.~Caianiello' dell'Universit\`{a} and Gruppo Collegato INFN, Salerno, Italy\\
$^{29}$ Dipartimento DISAT del Politecnico and Sezione INFN, Turin, Italy\\
$^{30}$ Dipartimento di Scienze MIFT, Universit\`{a} di Messina, Messina, Italy\\
$^{31}$ Dipartimento Interateneo di Fisica `M.~Merlin' and Sezione INFN, Bari, Italy\\
$^{32}$ European Organization for Nuclear Research (CERN), Geneva, Switzerland\\
$^{33}$ Faculty of Electrical Engineering, Mechanical Engineering and Naval Architecture, University of Split, Split, Croatia\\
$^{34}$ Faculty of Engineering and Science, Western Norway University of Applied Sciences, Bergen, Norway\\
$^{35}$ Faculty of Nuclear Sciences and Physical Engineering, Czech Technical University in Prague, Prague, Czech Republic\\
$^{36}$ Faculty of Physics, Sofia University, Sofia, Bulgaria\\
$^{37}$ Faculty of Science, P.J.~\v{S}af\'{a}rik University, Ko\v{s}ice, Slovak Republic\\
$^{38}$ Frankfurt Institute for Advanced Studies, Johann Wolfgang Goethe-Universit\"{a}t Frankfurt, Frankfurt, Germany\\
$^{39}$ Fudan University, Shanghai, China\\
$^{40}$ Gangneung-Wonju National University, Gangneung, Republic of Korea\\
$^{41}$ Gauhati University, Department of Physics, Guwahati, India\\
$^{42}$ Helmholtz-Institut f\"{u}r Strahlen- und Kernphysik, Rheinische Friedrich-Wilhelms-Universit\"{a}t Bonn, Bonn, Germany\\
$^{43}$ Helsinki Institute of Physics (HIP), Helsinki, Finland\\
$^{44}$ High Energy Physics Group,  Universidad Aut\'{o}noma de Puebla, Puebla, Mexico\\
$^{45}$ Horia Hulubei National Institute of Physics and Nuclear Engineering, Bucharest, Romania\\
$^{46}$ HUN-REN Wigner Research Centre for Physics, Budapest, Hungary\\
$^{47}$ Indian Institute of Technology Bombay (IIT), Mumbai, India\\
$^{48}$ Indian Institute of Technology Indore, Indore, India\\
$^{49}$ INFN, Laboratori Nazionali di Frascati, Frascati, Italy\\
$^{50}$ INFN, Sezione di Bari, Bari, Italy\\
$^{51}$ INFN, Sezione di Bologna, Bologna, Italy\\
$^{52}$ INFN, Sezione di Cagliari, Cagliari, Italy\\
$^{53}$ INFN, Sezione di Catania, Catania, Italy\\
$^{54}$ INFN, Sezione di Padova, Padova, Italy\\
$^{55}$ INFN, Sezione di Pavia, Pavia, Italy\\
$^{56}$ INFN, Sezione di Torino, Turin, Italy\\
$^{57}$ INFN, Sezione di Trieste, Trieste, Italy\\
$^{58}$ Inha University, Incheon, Republic of Korea\\
$^{59}$ Institute for Gravitational and Subatomic Physics (GRASP), Utrecht University/Nikhef, Utrecht, Netherlands\\
$^{60}$ Institute of Experimental Physics, Slovak Academy of Sciences, Ko\v{s}ice, Slovak Republic\\
$^{61}$ Institute of Physics, Homi Bhabha National Institute, Bhubaneswar, India\\
$^{62}$ Institute of Physics of the Czech Academy of Sciences, Prague, Czech Republic\\
$^{63}$ Institute of Space Science (ISS), Bucharest, Romania\\
$^{64}$ Institut f\"{u}r Kernphysik, Johann Wolfgang Goethe-Universit\"{a}t Frankfurt, Frankfurt, Germany\\
$^{65}$ Instituto de Ciencias Nucleares, Universidad Nacional Aut\'{o}noma de M\'{e}xico, Mexico City, Mexico\\
$^{66}$ Instituto de F\'{i}sica, Universidade Federal do Rio Grande do Sul (UFRGS), Porto Alegre, Brazil\\
$^{67}$ Instituto de F\'{\i}sica, Universidad Nacional Aut\'{o}noma de M\'{e}xico, Mexico City, Mexico\\
$^{68}$ iThemba LABS, National Research Foundation, Somerset West, South Africa\\
$^{69}$ Jeonbuk National University, Jeonju, Republic of Korea\\
$^{70}$ Johann-Wolfgang-Goethe Universit\"{a}t Frankfurt Institut f\"{u}r Informatik, Fachbereich Informatik und Mathematik, Frankfurt, Germany\\
$^{71}$ Korea Institute of Science and Technology Information, Daejeon, Republic of Korea\\
$^{72}$ KTO Karatay University, Konya, Turkey\\
$^{73}$ Laboratoire de Physique Subatomique et de Cosmologie, Universit\'{e} Grenoble-Alpes, CNRS-IN2P3, Grenoble, France\\
$^{74}$ Lawrence Berkeley National Laboratory, Berkeley, California, United States\\
$^{75}$ Lund University Department of Physics, Division of Particle Physics, Lund, Sweden\\
$^{76}$ Nagasaki Institute of Applied Science, Nagasaki, Japan\\
$^{77}$ Nara Women{'}s University (NWU), Nara, Japan\\
$^{78}$ National and Kapodistrian University of Athens, School of Science, Department of Physics , Athens, Greece\\
$^{79}$ National Centre for Nuclear Research, Warsaw, Poland\\
$^{80}$ National Institute of Science Education and Research, Homi Bhabha National Institute, Jatni, India\\
$^{81}$ National Nuclear Research Center, Baku, Azerbaijan\\
$^{82}$ National Research and Innovation Agency - BRIN, Jakarta, Indonesia\\
$^{83}$ Niels Bohr Institute, University of Copenhagen, Copenhagen, Denmark\\
$^{84}$ Nikhef, National institute for subatomic physics, Amsterdam, Netherlands\\
$^{85}$ Nuclear Physics Group, STFC Daresbury Laboratory, Daresbury, United Kingdom\\
$^{86}$ Nuclear Physics Institute of the Czech Academy of Sciences, Husinec-\v{R}e\v{z}, Czech Republic\\
$^{87}$ Oak Ridge National Laboratory, Oak Ridge, Tennessee, United States\\
$^{88}$ Ohio State University, Columbus, Ohio, United States\\
$^{89}$ Physics department, Faculty of science, University of Zagreb, Zagreb, Croatia\\
$^{90}$ Physics Department, Panjab University, Chandigarh, India\\
$^{91}$ Physics Department, University of Jammu, Jammu, India\\
$^{92}$ Physics Program and International Institute for Sustainability with Knotted Chiral Meta Matter (SKCM2), Hiroshima University, Hiroshima, Japan\\
$^{93}$ Physikalisches Institut, Eberhard-Karls-Universit\"{a}t T\"{u}bingen, T\"{u}bingen, Germany\\
$^{94}$ Physikalisches Institut, Ruprecht-Karls-Universit\"{a}t Heidelberg, Heidelberg, Germany\\
$^{95}$ Physik Department, Technische Universit\"{a}t M\"{u}nchen, Munich, Germany\\
$^{96}$ Politecnico di Bari and Sezione INFN, Bari, Italy\\
$^{97}$ Research Division and ExtreMe Matter Institute EMMI, GSI Helmholtzzentrum f\"ur Schwerionenforschung GmbH, Darmstadt, Germany\\
$^{98}$ Saga University, Saga, Japan\\
$^{99}$ Saha Institute of Nuclear Physics, Homi Bhabha National Institute, Kolkata, India\\
$^{100}$ School of Physics and Astronomy, University of Birmingham, Birmingham, United Kingdom\\
$^{101}$ Secci\'{o}n F\'{\i}sica, Departamento de Ciencias, Pontificia Universidad Cat\'{o}lica del Per\'{u}, Lima, Peru\\
$^{102}$ Stefan Meyer Institut f\"{u}r Subatomare Physik (SMI), Vienna, Austria\\
$^{103}$ SUBATECH, IMT Atlantique, Nantes Universit\'{e}, CNRS-IN2P3, Nantes, France\\
$^{104}$ Sungkyunkwan University, Suwon City, Republic of Korea\\
$^{105}$ Suranaree University of Technology, Nakhon Ratchasima, Thailand\\
$^{106}$ Technical University of Ko\v{s}ice, Ko\v{s}ice, Slovak Republic\\
$^{107}$ The Henryk Niewodniczanski Institute of Nuclear Physics, Polish Academy of Sciences, Cracow, Poland\\
$^{108}$ The University of Texas at Austin, Austin, Texas, United States\\
$^{109}$ Universidad Aut\'{o}noma de Sinaloa, Culiac\'{a}n, Mexico\\
$^{110}$ Universidade de S\~{a}o Paulo (USP), S\~{a}o Paulo, Brazil\\
$^{111}$ Universidade Estadual de Campinas (UNICAMP), Campinas, Brazil\\
$^{112}$ Universidade Federal do ABC, Santo Andre, Brazil\\
$^{113}$ Universitatea Nationala de Stiinta si Tehnologie Politehnica Bucuresti, Bucharest, Romania\\
$^{114}$ University of Cape Town, Cape Town, South Africa\\
$^{115}$ University of Derby, Derby, United Kingdom\\
$^{116}$ University of Houston, Houston, Texas, United States\\
$^{117}$ University of Jyv\"{a}skyl\"{a}, Jyv\"{a}skyl\"{a}, Finland\\
$^{118}$ University of Kansas, Lawrence, Kansas, United States\\
$^{119}$ University of Liverpool, Liverpool, United Kingdom\\
$^{120}$ University of Science and Technology of China, Hefei, China\\
$^{121}$ University of South-Eastern Norway, Kongsberg, Norway\\
$^{122}$ University of Tennessee, Knoxville, Tennessee, United States\\
$^{123}$ University of the Witwatersrand, Johannesburg, South Africa\\
$^{124}$ University of Tokyo, Tokyo, Japan\\
$^{125}$ University of Tsukuba, Tsukuba, Japan\\
$^{126}$ Universit\"{a}t M\"{u}nster, Institut f\"{u}r Kernphysik, M\"{u}nster, Germany\\
$^{127}$ Universit\'{e} Clermont Auvergne, CNRS/IN2P3, LPC, Clermont-Ferrand, France\\
$^{128}$ Universit\'{e} de Lyon, CNRS/IN2P3, Institut de Physique des 2 Infinis de Lyon, Lyon, France\\
$^{129}$ Universit\'{e} de Strasbourg, CNRS, IPHC UMR 7178, F-67000 Strasbourg, France, Strasbourg, France\\
$^{130}$ Universit\'{e} Paris-Saclay, Centre d'Etudes de Saclay (CEA), IRFU, D\'{e}partment de Physique Nucl\'{e}aire (DPhN), Saclay, France\\
$^{131}$ Universit\'{e}  Paris-Saclay, CNRS/IN2P3, IJCLab, Orsay, France\\
$^{132}$ Universit\`{a} degli Studi di Foggia, Foggia, Italy\\
$^{133}$ Universit\`{a} del Piemonte Orientale, Vercelli, Italy\\
$^{134}$ Universit\`{a} di Brescia, Brescia, Italy\\
$^{135}$ Variable Energy Cyclotron Centre, Homi Bhabha National Institute, Kolkata, India\\
$^{136}$ Warsaw University of Technology, Warsaw, Poland\\
$^{137}$ Wayne State University, Detroit, Michigan, United States\\
$^{138}$ Yale University, New Haven, Connecticut, United States\\
$^{139}$ Yonsei University, Seoul, Republic of Korea\\
$^{140}$  Zentrum  f\"{u}r Technologie und Transfer (ZTT), Worms, Germany\\
$^{141}$ Affiliated with an institute covered by a cooperation agreement with CERN\\
$^{142}$ Affiliated with an international laboratory covered by a cooperation agreement with CERN.\\

\end{flushleft} 